\documentclass[showpacs,preprintnumbers,amssymb,nofootinbib]{revtex4}
\usepackage{epsfig}
\usepackage{amsmath}
\usepackage{graphicx}
\usepackage{amssymb}
\usepackage{bm}
\usepackage{dcolumn}
\usepackage[svgnames]{xcolor}

\newcommand{\ba}{\begin{array}}
\newcommand{\ea}{\end{array}}
\def\beq{\begin{equation}}
\def\eeq{\end{equation}}
\def\bea{\begin{eqnarray}}
\def\eea{\end{eqnarray}}
\def\nn{\nonumber}
\def\sss{\scriptscriptstyle}

\def\roughly#1{\mathrel{\raise.3ex\hbox
{$#1$\kern-.75em\lower1ex\hbox{$\sim$}}}}

\def\sla#1{\raise.15ex\hbox{$/$}\kern-.57em #1}

\def\ket#1{\left| #1\right\rangle}
\def\ks{K_S}

\def\bd{B_d^0}
\def\bs{B_s^0}

\def\btos{{\bar b} \to {\bar s}}
\def\order{\lower 1.8ex \hbox{\LARGE\~{}}}

\newcommand{\lt}{\left}
\newcommand{\rt}{\right}

\def\bdbar{{\bar B}_d^0}
\def\bsbar{{\bar B}_s^0}

\def\btopik{B \to \pi K}
\def\fT{f_{\sss T}}
\def\fL{f_{\sss L}}
\def\fTfL{f_{\sss T}/f_{\sss L}}

\def\Bsdecay{\bsbar\to J/\psi \phi}

\def\bkll{{\bar B} \to {\bar K}^* \mu^+ \, \mu^-}
\def\Bsdecay{\bs \to J/\psi \phi}
\def\bscc{{\bar b} \to {\bar s} c {\bar c}}

\def\be {\begin{equation}}
\def\ee {\end{equation}}

\begin{document}
\preprint{UdeM-GPP-TH-11-202 }
\title{\boldmath $\bs (\bsbar) \to D^0_{CP} K {\bar K}$: Detecting and Discriminating New Physics in $\bs$-$\bsbar$ Mixing}

\author{Soumitra Nandi}
\affiliation{Physique des Particules, Universit\'e de
  Montr\'eal,\\ C.P. 6128, succ.\ centre-ville, Montr\'eal, QC, Canada
  H3C 3J7 }

\author{David London}
\affiliation{Physique des Particules, Universit\'e de
  Montr\'eal,\\ C.P. 6128, succ.\ centre-ville, Montr\'eal, QC, Canada
  H3C 3J7 }

\begin{abstract}
If the weak phase of $\bs$-$\bsbar$ mixing ($2\beta_s$) is found to be
significantly different from zero, this is a clear signal of new
physics (NP). However, if such a signal is found, we would like an
unambiguous determination of $2\beta_s$ in order to ascertain which NP
models could be responsible. In addition, in the presence of NP, the
width difference $\Delta\Gamma_s$ between the two $B_s$ mass
eigenstates can be positive or negative, and ideally this sign
ambiguity should be resolved experimentally.  Finally, in order to see
if the NP is contributing to $\Gamma^s_{12}$ in addition to
$M^s_{12}$, the precise measurement of $|\Gamma^s_{12}|$ is
crucial. In this paper, we consider several different methods of
measuring $\bs$-$\bsbar$ mixing using two- and three-body decays with
${\bar b} \to {\bar c} u {\bar s}$ and ${\bar b} \to {\bar u} c {\bar
  s}$ transitions.  We find that the most promising of these is a
time-dependent Dalitz-plot analysis of $\bs (\bsbar) \to D^0_{CP} K
{\bar K}$. With these decays, all of the above issues can be
addressed, and the measurement of the weak phase $\gamma$ is also
possible. We also note that, with all three-body decays it is possible
to resolve the sign ambiguity of $\Delta\Gamma_s$ even without
determining CP phase $\phi_s$.
\end{abstract}

\maketitle

\section{Introduction}

Over the past several years, a number of discrepancies with the
predictions of the standard model (SM) have been observed in $B$
decays, intriguingly all in $\btos$ transitions. Some examples are:
(i) in $\btopik$ decays, it is difficult to account for all the
experimental measurements within the SM \cite{kundu_nandi,piKupdate},
(ii) the values of the $\bd$-$\bdbar$ mixing phase $\sin 2\beta$
obtained from different penguin-dominated $\btos$ channels tend to be
systematically smaller than that obtained from $\bd\to J/\psi \ks$
\cite{btos}, (iii) the fractions of transversely- and
longitudinally-polarized decays in $B\to\phi K^*$ ($\fT$ and $\fL$,
respectively) are observed to be roughly equal \cite{phiK*}, in
contrast to the naive expectation that $\fTfL \ll 1$, (iv) the
differential forward-backward asymmetry of leptons in the exclusive
decay $\bkll$ is found to differ from the SM expectations in both the
low- and high-$q^2$ regions ($q^2$ is the dilepton invariant mass)
\cite{Belle,BaBar}.

In light of this, it is particularly important to study $\btos$
transitions and look for new-physics (NP) effects. Now, if NP is
present in $\Delta B = 1$ $\btos$ decays, it would be highly unnatural
for it not to also affect the $\Delta B=2$ transition, in particular
$\bs$-$\bsbar$ mixing. In order to see where NP can enter, we briefly
review the mixing. In the $B_s$ system, the mass eigenstates $B_{L}$
and $B_{H}$ ($L$ and $H$ indicate the light and heavy states,
respectively) are admixtures of the flavor eigenstates $\bs$ and
$\bsbar$:
\bea 
\ket{B_L} &=& p \ket{\bs} + q \ket{\bsbar} ~, \nn\\
\ket{B_H} &=& p \ket{\bs} - q \ket{\bsbar} ~, 
\eea 
with $|p|^2 + |q|^2 =1$.  As a result, the initial flavor eigenstates
oscillate into one another according to the Schr\"odinger equation
\bea 
i \frac{d}{dt} \left( \ba{c} \ket{\bs(t)} \\ \ket{\bsbar(t)} \ea \right) =
\left(M^s -i \frac{\Gamma^s}{2} \right) \left( \ba{c} \ket{\bs(t)}
\\ \ket{\bsbar(t)} \ea \right) ~,
\eea 
where $M=M^\dagger$ and $\Gamma=\Gamma^\dagger$ correspond
respectively to the dispersive and absorptive parts of the mass
matrix. The off-diagonal elements, $M^s_{12}=M_{21}^{s*}$ and
$\Gamma^s_{12}=\Gamma_{21}^{s*}$, are generated by $\bs$-$\bsbar$
mixing. We define
\bea 
\Gamma_s \equiv \frac{\Gamma_H
  + \Gamma_L}{2}, \quad \Delta M_s \equiv   M_H - M_L, \quad \Delta \Gamma_s \equiv   \Gamma_L -
\Gamma_H ~.
\eea 
Expanding the mass eigenstates, we find, to a very good approximation
\cite{Dunietz:2000},
\bea 
\Delta M_s &=& 2 |M_{12}^s| ~, \nn\\
 \Delta \Gamma_s &=&  2 |\Gamma_{12}^s| \cos \phi_s  ~, \nn \\
 \frac{q}{p} &=&  e^{-2i\beta_s}\left[1-\frac{a}{2}\right] ~,
\label{eq:mix_para}
\eea 
where $\phi_s \equiv \arg(-M_{12}^s/\Gamma_{12}^s)$ is the CP phase in
$\Delta B=2$ transitions, and $2\beta_s = \arg(M_{12}^s)$. In
Eq.~(\ref{eq:mix_para}) the small expansion parameter $a$ is given by
\be
a = \frac{\Gamma^s_{12}}{M^s_{12}}\,\sin\phi_s ~.
\ee
This is expected to be $\ll 1$, and hence can be neglected in the
definition of $q/p$.  It is also important to note that the sign of
$\Delta\Gamma_s$ is equal to the sign of $\cos\phi_s$, and in the case
where there is no NP in $\Gamma_{12}^s$, the CP phase $\phi_s = - 2
\beta_s $.

The precise measurement of $\Delta M_s$ determines $|M^s_{12}|$
\cite{bsmixing}.  However, because of hadronic uncertainties, the SM
prediction for $\Delta M_s$ is not very precise -- in
Ref.~\cite{lenz_uli}, it is noted that the theoretical uncertainties
still allow new-physics contributions to $|M^s_{12}|$ of order 20\%.
In addition, $\Gamma^s_{12}$ can be calculated from the absorptive
part of the $\bs$-$\bsbar$ mixing box diagram, leading to $\Delta
\Gamma_s$. Unlike the $B_d$ system, where $\Delta\Gamma_d$ is
negligibly small, in the $B_s$ system $\Delta \Gamma_s$ is expected to
be reasonably large, which leads to certain advantages for the search
for CP-violating effects in the $B_s$ system over that of $B_d$
system. The updated SM predictions of the width difference and the CP
phase $\phi_s$ are given by \cite{lenz_uli}
\bea
\Delta\Gamma^{SM}_s &\simeq& 2 |\Gamma^s_{12}|= 0.087 \pm 0.021 ~~ ps^{-1}, ~ \nonumber \\
\phi_s &\approx& 0.22^{\circ} ~.
\label{DGSSM}
\eea

Although the SM predictions for $\Delta M_s$ and $\Delta \Gamma_s$ are
not precise, the SM does predict that $2\beta_s \simeq 0$, which makes
it a good observable to use in the search for NP. Consider first the
case where the NP contributes only to $M^s_{12}$.  If one measures a
value for $2\beta_s$ that is significantly different from zero, this
will indicate NP in $\bs$-$\bsbar$ mixing (in $M^s_{12}$).  However,
to cover all bases, one more step must be done. Suppose that NP is
present, but it produces $2\beta_s = 180^\circ$. Now indirect CP
violation, which measures $\sin 2\beta_s$, will not give a signal. But
one can still detect the NP by measuring the sign of $\Delta\Gamma_s$
-- in the SM, $\Delta \Gamma_s > 0$, while it is $<0$ if $2\beta_s =
180^\circ$. Also, even if NP is discovered through indirect CP
violation (i.e.\ $\sin 2\beta_s \ne 0$), this only determines
$2\beta_s$ up to a twofold discrete ambiguity. Since the sign of
$\cos2\beta_s$ can be determined by the sign of $\Delta\Gamma_s$, the
knowledge of this sign is one possibility to remove this discrete
ambiguity. Alternatively, one may try to find a method which allows a
direct determination of $2\beta_s$ without any ambiguity. Now suppose
that the NP also contributes to $\Gamma_{12}^s$. Since there is NP in
$M^s_{12}$, its presence can be detected as above. But now the twofold
ambiguity in $2\beta_s$ cannot be removed from the knowledge of the
sign of $\Delta\Gamma_s$, since this only determines the sign of
$\cos\phi_s$ (and not $\cos2\beta_s$). Thus, for the case where there
is NP in $\Gamma_{12}^s$, one must find another way to remove the
twofold ambiguity in $2\beta_s$.

The CDF \cite{CDF} and D\O\ \cite{D0} collaborations have measured the
CP asymmetry in $\Bsdecay$, and found a hint for indirect CP
violation. In general, this result is interpreted as evidence for a
nonzero value of $2\beta^{\psi\phi}_s$, and the contributions of various NP
models to the $B_s$ mixing phase have been explored \cite{RPV,Z'FCNC,
  2HDM, SUSY, littleHiggs,fourGen, Bspapers}.  It has also been
pointed out that NP in the decay $\bscc$ could also play an important
role \cite{NPdecay}. Recently CDF and D\O\ updated their measurements
of the CP-violating phase. The 68\% C.L. allowed ranges are
\cite{cdfnew,d0new}
\begin{align}
2\beta^{\psi\phi}_s \,&\in \, \lt[2.3^\circ, 59.6^\circ
  \rt]\cup\lt[123.8^\circ,177.6^\circ \rt] ~, &
{\hbox{CDF}} ~,  \nn \\
                   & \in \, \lt[ 9.7^\circ , 52.1^\circ
  \rt]\cup\lt[127.9^\circ , 170.3^\circ \rt] ~, &
{\hbox{D\O}} ~.
\label{cdfphi}
\end{align}
Most of the values of $2\beta^{\psi\phi}_s$ here suggest NP. $2\beta^{\psi\phi}_s$ is
obtained with the twofold ambiguity $2\beta^{\psi\phi}_s \leftrightarrow \pi -
2\beta^{\psi\phi}_s$, and at present there is no preference for either of the two
solutions.  As mentioned above, the possibility of NP in the decay
$\bscc$ cannot be ruled out, so that the phase $2\beta^{\psi\phi}_s$ extracted
from $\Bsdecay$ should not necessarily be taken as purely a mixing
phase.  It is therefore worthwhile to look for a process in which NP
in the decay can essentially be neglected, and which permits the
determination of $2\beta_s$ without any ambiguity. If the measured
value of $2\beta_s$ is found to be significantly different from that
in $\Bsdecay$, it will be clear signal of NP in $\bscc$.

In addition, the D\O\ Collaboration recently found a large CP
asymmetry in the like-sign dimuon signal, which they attribute
primarily to $a^s_{SL}$, the semileptonic CP asymmetry in $\bs \to X_s
\mu \nu$ \cite{d0dimuonprd,d0dimuonprl}.  Now, the D\O\ result is less
than 2$\sigma$ away from zero and consequently to an excellent
approximation also about 2$\sigma$ away from the SM prediction
($a^{s,SM}_{SL}\approx 2 \times 10^{-5}$) \cite{lenz_uli}. Still, NP
in $\bs$-$\bsbar$ mixing can explain the result (for example, see
Ref.~\cite{dimuon_mixing}).  However, if one wishes to reproduce the
central value of $a^s_{SL}$, one requires NP specifically in
$\Gamma^s_{12}$ \cite{dkn2,bauer}. There are NP models that can
contribute to $\Gamma^s_{12}$ through the decay $b\to s \tau^+ \tau^-$
\cite{dkn1,NPGamma12}, and a significant enhancement of its magnitude
over that of the SM \cite{lenz_uli} is possible.  Furthermore, the
possibility of NP effects in $\Gamma^s_{12}$ through the decay $\bscc$
cannot be ruled out \cite{dkn1,bauer}.

We therefore see that there are some hints of NP in the $B_s$ system,
but nothing definitive yet. Thus, it is important to look for
additional methods of probing NP in $\bs$-$\bsbar$ mixing. Ideally,
the new method(s) would allow an unambiguous determination of the
mixing phase $2\beta_s$. Also useful are methods which remove the sign
ambiguity in $\Delta\Gamma_s$ even without providing any direct
information on the CP phases $\phi_s$ or $2\beta_s$. Finally, if NP is
present in the mixing, we would like to know if it contributes to
$\Gamma^s_{12}$ in addition to $M^s_{12}$. Hence, along with the
removal of the sign ambiguity in $\Delta\Gamma_s$, independent and
unbiased measurements of $|\Gamma^s_{12}|$ and $\phi_s$ are essential.

Several years ago, the two-body decays $\bs ({\bsbar}) \to D^{\pm}_s
K^{\mp}, D^{*\pm}_s K^{\mp}, ...$ were examined with the idea of
extracting weak phases \cite{fleischerdsk}. Because the final state is
accessible to both $\bs$ and $\bsbar$ mesons, a mixing-induced
indirect CP asymmetry occurs. Using this, and assuming that $\Delta
\Gamma_s$ is sizeable, the conclusion of Ref.~\cite{fleischerdsk} is
that one can measure the phase $2\beta_s + \gamma$ with a twofold
discrete ambiguity, and that this ambiguity can be removed if
factorization is assumed. However, if there is NP in $\bs$-$\bsbar$
mixing, $\Delta \Gamma_s < 0$ is allowed as well. This implies that,
in fact, $2\beta_s + \gamma$ can be obtained with a fourfold discrete
ambiguity (or twofold if factorization is assumed).

In Ref.~\cite{nandi_uli} it was shown that the sign ambiguity in
$\Delta\Gamma_s$ can be removed using $\bs\to D_s^{\pm} K^{\mp}$
decays. Although the method does not allow a direct determination of
the phase $2\beta_s$, it does discriminate between the two solutions
with $\cos2\beta_s > 0$ and $\cos2\beta_s < 0$, which then determines
the sign of $\Delta\Gamma_s$. However, the method is based on several
assumptions: (i) the weak phase $\gamma$ is taken from the $B$-factory
measurements, (ii) factorization is assumed, i.e.\ the strong phase is
taken to be $\simeq 0$, and (iii) the SM-predicted value of
$\Gamma^s_{12}$ has been used in the analysis.

In 1991, the decays $\bs ({\bsbar}) \to D^0_{CP} \phi$, where
$D^0_{CP}$ is a neutral $D$-meson CP-eigenstate, were proposed to
extract the CKM angle $\gamma$ with a twofold ambiguity
\cite{Gronau:1990ra,fleischerdphi}. However, these methods assumed that
the phase $2\beta_s$ is approximately zero (or known). The current
experimental data [see Eq.~(\ref{cdfphi})] is not completely in favor
of this assumption -- there is the possibility that $2\beta_s$ can be 
significantly different from zero. In addition, at present $2\beta_s$
is measured with a twofold ambiguity, which adds a further discrete
ambiguity to the determination of $\gamma$.

We therefore see that previous analyses of two-body $B$ decays only
partially probe NP in $\bs$-$\bsbar$ mixing -- $2\beta_s$ is, in
general, not determined unambiguously, the sign ambiguity in
$\Delta\Gamma_s$ is generally unresolved, and the possibility of NP
affecting $\Gamma^s_{12}$ has not been considered.  In this paper we
go beyond the previous analyses to explore all of these issues.

In Sec.~\ref{2body} we review the two-body decays. In particular, in
Sec.~\ref{twobodydphi}, we update the analysis of $\bs ({\bsbar}) \to
D^0_{CP} \phi$, considering both $\Delta\Gamma_s>0$ and
$\Delta\Gamma_s <0$. In Sec.~\ref{threebody}, we present the
Dalitz-plot analyses of three-body decays. In particular, in
Sec.~\ref{threebodydspi}, we focus on $\bs ({\bsbar}) \to D^{\pm}_s
K^{\mp} \pi^0, D^{\pm}_s \pi^{\mp} K^0, ...$, using the interference
between the different intermediate resonant decays to provide
additional information.  And in Sec.~\ref{threebodydkk}, we show that
a much greater improvement can be obtained by performing a
time-dependent Dalitz-plot analysis of the decay $\bs ({\bsbar})\to
D^0_{CP} K {\bar K}$. Finally, in Sec.~\ref{gamma12exp}, we present a
possible way to determine $\Delta\Gamma_s$, or equivalently
$|\Gamma^s_{12}|$ and $\phi_s$, using three-body decays.  We conclude
in Sec.~\ref{conclusions}.

\section{Two-Body Decays}
\label{2body}

\subsection{\boldmath $\bs (\bsbar) \to f, {\bar f}$}

Consider a final state $f$, not necessarily a CP eigenstate, to which
both $\bs$ and $\bsbar$ can decay.  In the presence of $\bs$-$\bsbar$
mixing, the time-dependent decay rates are given by
\cite{Dunietz}
\bea
\Gamma(\bs(t) \to f) &\! \sim \!& \frac12 e^{-\Gamma_s t} \Big\{ (|A_f|^2 + |{\bar A}_f|^2) \cosh(\Delta \Gamma_s t /2)
                                       + (|A_f|^2 - |{\bar A}_f|^2) \cos \Delta m_s t  \nn\\
&&  -~2 \sinh(\Delta \Gamma_s t /2) \, {\rm Re} \left[ \frac{q}{p} A_f^* {\bar A}_f  \right]
-2 \sin \Delta m_s t \, {\rm Im} \left[ \frac{q}{p} A_f^* {\bar A}_f \right] \Big\} ~, \nn\\
\Gamma(\bsbar(t) \to f) &\! \sim \!& \frac12 e^{-\Gamma_s t} \Big\{ (|A_f|^2 + |{\bar A}_f|^2) \cosh(\Delta \Gamma_s t /2)
                                       - (|A_f|^2 - |{\bar A}_f|^2) \cos \Delta m_s t  \nn\\
&&  -~2 \sinh(\Delta \Gamma_s t /2) \, {\rm Re} \left[ \frac{q}{p} A_f^* {\bar A}_f  \right]
+2 \sin \Delta m_s t \, {\rm Im} \left[ \frac{q}{p} A_f^* {\bar A}_f \right] \Big\} ~,
\label{timedeprates}
\eea
where $A_f \equiv A(\bs\to f)$, ${\bar A}_f \equiv A(\bsbar\to f )$, and $q/p = e^{-2i\beta_s}$.  This yields
\bea
\Gamma(\bs(t) \to f) - \Gamma(\bsbar(t) \to f) & \sim & (|A_f|^2 + |{\bar A}_f|^2) e^{-\Gamma_s t} \left[ 
C \cos \Delta m_s t - S \sin \Delta m_s t \right] ~, \nn\\
\Gamma(\bs(t) \to f) + \Gamma(\bsbar(t) \to f) & \sim & (|A_f|^2 + |{\bar A}_f|^2) e^{-\Gamma_s t} \left[ 
\cosh(\Delta \Gamma_s t /2) - {\cal A}_{\Delta\Gamma} \sinh(\Delta \Gamma_s t /2) \right] ~, 
\label{indirectCPA1}
\eea
where
\beq
C \equiv \frac{1 - |\lambda|^2}{1 + |\lambda|^2} ~,~~
S \equiv \frac {2 \, {\rm Im} \lambda} {1 + |\lambda|^2} ~,~~
{\cal A}_{\Delta\Gamma} \equiv \frac {2 \, {\rm Re} \lambda} {1 + |\lambda|^2} ~,~~
\lambda \equiv \frac{q}{p} \frac{{\bar A}_f}{A_f} ~.
\label{para:cp}
\eeq
The idea is that, by fitting the data corresponding to the difference
(``tagged'') and sum (``untagged'') of decay rates to the four
time-dependent functions given on the right-hand side of the equations
in Eq.~(\ref{indirectCPA1}), the coefficients of these functions can
be obtained, from which $C$, $S$ and ${\cal A}_{\Delta\Gamma}$ can be
derived.  However, there is a complication -- in the presence of NP in
$\Delta B=2$ transitions, $\Delta \Gamma_s$ is unknown (though it is
assumed to be reasonably large). Therefore, for the untagged
combination, both $\Delta \Gamma_s$ and ${\cal A}_{\Delta\Gamma}$ must
be found in the fit. Still, though this will determine $|\Delta
\Gamma_s|$, its sign will remain unknown. The reason is that only the
function $\sinh(\Delta \Gamma_s t /2)$ is sensitive to the sign of
$\Delta \Gamma_s$, and it is multiplied by ${\cal
  A}_{\Delta\Gamma}$. Thus, any change in the sign of $\Delta
\Gamma_s$ can be compensated for by changing the sign of ${\cal
  A}_{\Delta\Gamma}$. The bottom line is that any analysis which uses
${\cal A}_{\Delta\Gamma}$ will have a discrete ambiguity due to the
unknown sign of $\Delta \Gamma_s$.

Similarly,
\bea
\Gamma(\bs(t) \to {\bar f}) &\! \sim \!& \frac12 e^{-\Gamma_s t} \Big\{ (|A_{\bar f}|^2 + |{\bar A}_{\bar f}|^2) \cosh(\Delta \Gamma_s t /2)
                                       + (|A_{\bar f}|^2 - |{\bar A}_{\bar f}|^2) \cos \Delta m_s t  \nn\\
&&  -~2 \sinh(\Delta \Gamma_s t /2) \, {\rm Re} \left[ \frac{p}{q} {\bar A}_{\bar f}^* A_{\bar f}  \right]
+2 \sin \Delta m_s t \, {\rm Im} \left[ \frac{p}{q} {\bar A}_{\bar f}^* A_{\bar f} \right] \Big\} ~, \nn\\
\Gamma(\bsbar(t) \to {\bar f}) &\! \sim \!& \frac12 e^{-\Gamma_s t} \Big\{ (|A_{\bar f}|^2 + |{\bar A}_{\bar f}|^2) \cosh(\Delta \Gamma_s t /2)
                                       - (|A_{\bar f}|^2 - |{\bar A}_{\bar f}|^2) \cos \Delta m_s t  \nn\\
&&  -~2 \sinh(\Delta \Gamma_s t /2) \, {\rm Re} \left[ \frac{p}{q} {\bar A}_{\bar f}^* A_{\bar f}  \right]
-2 \sin \Delta m_s t \, {\rm Im} \left[ \frac{p}{q} {\bar A}_{\bar f}^* A_{\bar f} \right] \Big\} ~,
\eea
where $A_{\bar f} \equiv A(\bs\to {\bar f})$ and ${\bar A}_{\bar f}
\equiv A(\bsbar\to {\bar f})$.  Then
\beq
\frac{\Gamma(\bs(t) \to {\bar f}) - \Gamma(\bsbar(t) \to {\bar f})}
{\Gamma(\bs(t) \to {\bar f}) + \Gamma(\bsbar(t) \to {\bar f})} =
\frac{ {\bar C} \cos \Delta m_s t + {\bar S} \sin \Delta m_s t}
{ \cosh(\Delta \Gamma_s t /2) - {\bar{\cal A}}_{\Delta\Gamma} \sinh(\Delta \Gamma_s t /2)} ~,
\label{indirectCPA2}
\eeq
where
\beq
{\bar C} \equiv \frac{1 - |{\bar\lambda}|^2}{1 + |{\bar\lambda}|^2} ~,~~
{\bar S} \equiv \frac {2 \, {\rm Im} {\bar\lambda}} {1 + |{\bar\lambda}|^2} ~,~~
{\bar{\cal A}}_{\Delta\Gamma} \equiv \frac {2 \, {\rm Re} {\bar\lambda}} {1 + |{\bar\lambda}|^2} ~,~~
{\bar\lambda} \equiv \frac{p}{q} \frac{A_{\bar f}}{{\bar A}_{\bar f}} ~.
\label{para:cpa}
\eeq

\subsection{\boldmath $\bs (\bsbar) \to D^{\pm}_s K^{\mp}$} 

Consider the decay $\bs ({\bsbar}) \to PP$ ($P$ is a pseudoscalar), in
which the final state contains a single $c$ quark\footnote{Much of the
  discussion in this subsection can be found in
  Ref.~\cite{fleischerdsk}, except that here NP in $\Delta \Gamma_s$
  is considered.}. Excluding those final states involving $\eta$'s,
there are only two decays in which the $\bs$ and $\bsbar$ amplitudes
are of comparable size: $\bs ({\bsbar}) \to D_s^- K^+$ and $\bs
({\bsbar}) \to D_s^+ K^-$. The $\bs$ decays are mediated by
color-allowed tree-level transitions ${\bar b} \to {\bar c} u {\bar
  s}$ and ${\bar b} \to {\bar u} c {\bar s}$. Within the SM, the
amplitudes take the form\footnote{In Ref.~\cite{bauer}, it is shown
  that NP in the decays ${\bar b} \to {\bar c} u {\bar s}$ and ${\bar
    b} \to {\bar u} c {\bar s}$ is strongly constrained. Such NP
  contributions are therefore neglected throughout this paper.} (there
is a minus sign associated with the ${\bar u}$ quark)
\bea
A(\bs\to D_s^- K^+) &=& T', \hskip 50pt A(\bs\to D_s^+ K^-) = -{\tilde T}' e^{i\gamma} ~, \nn\\
A(\bsbar\to D_s^- K^+) &=& {\tilde T}' e^{-i\gamma} ~, \hskip 28pt A(\bsbar\to D_s^+ K^-) = -T' ~.
\label{DsKamps}
\eea
We have explicitly written the weak-phase dependence, while the
diagrams contain strong phases.  The magnitudes of the
Cabbibo-Kobayashi-Maskawa (CKM) matrix elements $|V_{cb}^* V_{us}|$
and $|V_{ub}^* V_{cs}|$ have been absorbed into the diagrams $T'$ and
${\tilde T}'$, respectively. (As this is a $\btos$ transition, the
diagrams are written with primes.)

Using the amplitudes of Eq.~(\ref{DsKamps}), one obtains [see
  Eqs.~(\ref{para:cp}) and (\ref{para:cpa})]
\bea
C &=& \frac{1 - |\lambda|^2}{1 + |\lambda|^2} ~,~~ S = -\frac{2|\lambda|}{1 + |\lambda|^2} \sin(2\beta_s + \gamma - \delta) ~,~~
{\cal A}_{\Delta\Gamma} = \frac{2|\lambda|}{1 + |\lambda|^2} \cos(2\beta_s + \gamma - \delta) ~, \nn\\
{\bar C} &=& \frac{1 - |\lambda|^2}{1 + |\lambda|^2} ~,~~ {\bar S} = \frac{2|\lambda|}{1 + |\lambda|^2} \sin(2\beta_s + \gamma + \delta) ~,~~~~
{\bar{\cal A}}_{\Delta\Gamma} = \frac{2|\lambda|}{1 + |\lambda|^2} \cos(2\beta_s + \gamma + \delta) ~,
\label{2bodyresults}
\eea
where $|\lambda| = {\tilde T}' / T'$ (defined to be positive) and
$\delta$ is the strong-phase difference between ${\tilde T}'$ and
$T'$. $|\lambda|$ can be obtained from the measurement of $C$. Using
this, $S$ and ${\cal A}_{\Delta\Gamma}$ give $\sin(2\beta_s + \gamma -
\delta)$ and $\cos(2\beta_s + \gamma - \delta)$, respectively. Thus,
one obtains $2\beta_s + \gamma - \delta$ with no discrete
ambiguity. Similarly, $2\beta_s + \gamma + \delta$ can be obtained
with no discrete ambiguity from {$\bar S$} and ${\bar{\cal
    A}}_{\Delta\Gamma}$. These can be combined to give the phases
($2\beta_s + \gamma$, $\delta$) with a twofold ambiguity [($2\beta_s +
  \gamma$, $\delta$) or ($2\beta_s + \gamma + \pi$, $\delta + \pi$)].
This discrete ambiguity can be removed if one assumes factorization,
which predicts $\delta$ to be near 0.

In fact, this is not quite correct. As discussed below
Eq.~(\ref{para:cp}), in the presence of NP in $\bs$-$\bsbar$ mixing
there is an additional discrete ambiguity due to the unknown sign of
$\Delta \Gamma_s$. Thus, the two-body decays $\bs (\bsbar) \to
D^{\pm}_s K^{\mp}$ permit the extraction of $2\beta_s + \gamma$ with a
fourfold ambiguity (or twofold if factorization is assumed).

Now, the value of $\gamma$ can be taken from the independent
measurements at the $B$-factories. One then obtains $2\beta_s$ with a
fourfold ambiguity. Alternatively, since $\gamma$ has not been
measured in $B_s$ decays, it can be kept with the aim of determining
its value independently (this was the original purpose of
Ref.~\cite{fleischerdsk}.) We adopt this latter approach in much of
the paper.

We therefore see that this method permits the extraction of $2\beta_s
+ \gamma$ with a fourfold ambiguity (or twofold if factorization is
assumed). It does not resolve the sign ambiguity in $\Delta\Gamma_s$,
and says nothing about the possibility of NP affecting
$\Gamma^s_{12}$. In order to address these remaining points, it is
necessary to examine other methods. A first step involves the decays
$\bs (\bsbar) \to D^0_{CP} \phi$, discussed in the next subsection.

\subsection{\boldmath $\bs (\bsbar) \to D^0_{CP} \phi$}
\label{twobodydphi}

Another pair of decays to which the method of the previous subsection
can be applied is $\bs (\bsbar) \to D^0 \phi, {\bar D^0} \phi$.  Here
the decays are mediated by color-suppressed tree-level
transitions. The amplitudes (of comparable size) are given by 
\bea
A(\bs\to  D^0 \phi) &=& -C^{\phi}_1 e^{i\gamma} ~, \hskip 40pt A(\bs\to {\bar D^0} \phi) =  C^{\phi}_2  ~, \nn\\
A(\bsbar\to {\bar D^0} \phi) &=&  C^{\phi}_1 e^{-i\gamma} ~, \hskip 30pt A(\bsbar\to D^0 \phi) = -C^{\phi}_2 ~.
\label{amps}
\eea
By measuring the time dependence of the decays, one can obtain $S$,
$\bar S$, $A_{\Delta\Gamma}$ and ${\bar A}_{\Delta\Gamma}$ as given in
Eqs.~(\ref{para:cp}) and (\ref{para:cpa}). Using these observables we
define
\bea
\sin(2\beta_s + \gamma + \delta_{\phi}) &=& -\frac{1 + |\lambda|^2}{2 |\lambda|} S \equiv S_{D} ~, \hskip 30pt 
\sin(2\beta_s + \gamma - \delta_{\phi}) = \frac{1 + |\lambda|^2}{2 |\lambda|} {\bar S} \equiv {\bar S}_{D} ~, \nn \\
\cos(2\beta_s + \gamma + \delta_{\phi}) &=& \frac{1 + |\lambda|^2}{2 |\lambda|} A_{\Delta\Gamma} \equiv A^D_{\Delta\Gamma} ~, \hskip 10pt 
\cos(2\beta_s + \gamma - \delta_{\phi}) = \frac{1 + |\lambda|^2}{2 |\lambda|} {\bar A}_{\Delta\Gamma} \equiv {\bar A^D}_{\Delta\Gamma} ~,
\label{cpasydphi}
\eea
with $\delta_{\phi} = \arg(C^{\phi}_1 / C^{\phi}_2)$. The method
of the previous subsection then allows us to obtain $2\beta_s +
\gamma$ with a twofold ambiguity (for the moment, we put aside the
ambiguity due to the sign of $\Delta \Gamma_s$).

The advantage of these decays is that there is a third decay which is
related: $\bs (\bsbar) \to D^0_{CP} \phi$, where $D^0_{CP}$ is a CP
eigenstate (either CP-odd or CP-even). In our analysis we consider
$D^0_{CP}$ as the CP-even superposition $(D^0 + {\bar D^0})/\sqrt{2}$.
The amplitudes for the decays are then given by
\bea
\sqrt{2} A(\bs\to D^0_{CP} \phi) &=& -C^{\phi}_1 e^{i\gamma} + C^{\phi}_2  ~, \nn\\
\sqrt{2} A(\bsbar\to D^0_{CP}\phi) &=&  C^{\phi}_1 e^{-i\gamma} - C^{\phi}_2 ~.
\label{ampdphi}
\eea
By measuring the time-dependent decay amplitudes of $\bs (\bsbar) \to
D \phi$ ($D = D^0,\bar D^0, D^0_{CP}$), one can extract the magnitudes
$|C^{\phi}_1|$, $|C^{\phi}_2|$, $|A_{D_{CP}}| = |A(\bs\to D^0_{CP}
\phi)|$ and $|{\bar A}_{D_{CP}}| = |A(\bsbar\to D^0_{CP} \phi)|$ (they
are combinations of the overall normalizations and the $C$ parameters
[Eq.~(\ref{para:cp})]).

Using the first equation of Eq.~(\ref{ampdphi}), we define
\be
\cos(\gamma + \delta_{\phi}) = \frac{2 |A_{D_{CP}}|^2 - |C^{\phi}_1|^2 - |C^{\phi}_2|^2 }{2 |C^{\phi}_1| |C^{\phi}_2|} \equiv \Sigma^+ ~.
\label{trang1}
\ee
Similarly, from the second equation of Eq.~(\ref{ampdphi}), we get
\be
\cos(\gamma - \delta_{\phi}) = \frac{2 |{\bar A}_{D_{CP}}|^2 - |C^{\phi}_1|^2 - |C^{\phi}_2|^2 }{2 |C^{\phi}_1| |C^{\phi}_2| } \equiv \Sigma^- ~.
\label{trang2}
\ee
Therefore, in the case of the $\bs (\bsbar) \to D \phi$ decays, we
have two more observables, $\Sigma^+$ and $\Sigma^-$. Combining
Eqs.~(\ref{cpasydphi}), (\ref{trang1}) and (\ref{trang2}), it is
straightforward to find expressions for $\sin2\beta_s$,
$\cos2\beta_s$, $\sin(2\beta_s + 2\gamma)$ and $\cos(2\beta_s +
2\gamma)$ in terms of the above observables:
\bea
\sin2\beta_s &=& \frac{ S_{D}^2 - {\bar S}_{D}^2 + {\Sigma^+}^2 - {\Sigma^-}^2 } { 2 (S_{D} \Sigma^+ - {\bar S}_{D} \Sigma^-)}~, \hskip 10pt 
\sin(2\beta_s+ 2\gamma) = \frac{ S_{D}^2 - {\bar S}_{D}^2 - {\Sigma^+}^2 + {\Sigma^-}^2 } { 2 (S_{D} \Sigma^- - {\bar S}_{D} \Sigma^+)}~, \nn \\
\cos2\beta_s &=& \frac{ S_{D}^2 - {\bar S}_{D}^2 - {\Sigma^+}^2 + {\Sigma^-}^2 } { 2 ({\bar A}^D_{\Delta\Gamma} \Sigma^- - A^D_{\Delta\Gamma} \Sigma^+)}~, 
\hskip 10pt
\cos(2\beta_s+ 2\gamma) = \frac{ S_{D}^2 - {\bar S}_{D}^2 + {\Sigma^+}^2 - {\Sigma^-}^2 } { 2 ({\bar A}^D_{\Delta\Gamma} \Sigma^+ - 
A^D_{\Delta\Gamma} \Sigma^-)}~,
\label{Dphiresults}
\eea
with
\be
S_{D}^2 - {\bar S}_{D}^2 = - (A^{D}_{\Delta\Gamma})^2 + ({\bar A}^{D}_{\Delta\Gamma})^2. 
\ee

Many years ago, $\bs (\bsbar) \to D \phi$ decays were studied
\cite{Gronau:1990ra}, but without the dependence on $\Delta
\Gamma_s$. It was found that $\sin2\beta_s$ and $\sin(2\beta_s + 2
\gamma)$ could be obtained, which correspond to determining $2\beta_s$
with a twofold ambiguity and $2\gamma$ with a fourfold ambiguity.  In
the present case, the dependence on $\Delta\Gamma_s$ is included. This
allows us to obtain $A^{D}_{\Delta\Gamma}$ and ${\bar
  A}^{D}_{\Delta\Gamma}$, which then permits us to measure
$\cos2\beta_s$ and $\cos(2\beta_s + 2 \gamma)$, in addition to
$\sin2\beta_s$ and $\sin(2\beta_s + 2 \gamma)$
[Eq.~(\ref{Dphiresults})]. These measurements allow an unambiguous
determination of $2\beta_s$ and $2\gamma$. We therefore see that a
nonzero $\Delta \Gamma_s$ helps quite a bit in determining the weak
phases.  As has been discussed above, the sign of $\Delta\Gamma_s$ is
not known, which implies that $A^{D}_{\Delta\Gamma}$ and ${\bar
  A}^{D}_{\Delta\Gamma}$ also have a sign ambiguity. This means that,
in fact, $2\beta_s$ and $\gamma$ are determined up to a twofold and
fourfold\footnote{Using Eq.~(\ref{Dphiresults}), we can determine
  $\cos2\gamma$ without any ambiguity, whereas, due to the unknown
  sign of $A^D_{\Delta\Gamma}$ or ${\bar A}^D_{\Delta\Gamma}$,
  $\sin2\gamma$ can be determined only with a twofold
  ambiguity. Combining these two results, $2\gamma$ can therefore be
  determined with a twofold ambiguity (or $\gamma$ with a fourfold
  ambiguity).} ambiguity, respectively. Therefore, once we are able to
fix the sign of $\Delta\Gamma_s$, the $\bs (\bsbar) \to D \phi$ decays
might be considered as an alternative mode to probe simultaneously
$\gamma$ and $2\beta_s$.

We therefore see that two-body ${\bar b} \to {\bar c} u {\bar
  s}$/${\bar b} \to {\bar u} c {\bar s}$ decays do not provide
sufficient information to measure the CP phases $2\beta_s$ and
$2\gamma$ in an unambiguous manner. In the next section we show that
there are several ways to improve upon the two-body decay methods by
using a Dalitz-plot analysis of the corresponding three-body decays.

\section{Three-Body Decays}
\label{threebody}

\subsection{\boldmath $\bs (\bsbar) \to f, {\bar f}$}

In recent years, it has been shown that one can get useful information
from three-body $B$ decays. For instance, time-integrated Dalitz-plot
analyses of $\bs \to K \pi \pi$ and $\bs \to \pi K {\bar K}$ decays
have been proposed as a probe of $\gamma$ \cite{Ciuchini:2006st}. And
various tests of the SM, as well as the extraction of weak phases,
have been examined in the context of $B \to K \pi \pi$, $B \to K {\bar
  K} K$, $B \to \pi {\bar K} K$ and $B \to \pi \pi \pi$ decays
\cite{nicmax}. 

In the previous section we discussed two-body ${\bar b} \to {\bar c} u
{\bar s}$/${\bar b} \to {\bar u} c {\bar s}$ decays; in this section
we examine the corresponding three-body decays. In $\bs ({\bsbar}) \to
PPP$ decays which receive a tree contribution, there are 5 final-state
($f, {\bar f}$) pairs: ($D_s^- K^+ \pi^0$, $D_s^+ K^- \pi^0$), ($D_s^-
K^0 \pi^+$, $D_s^+ {\bar K}^0 \pi^-$), ($D^- K^+ {\bar K}^0$, $D^+ K^0
K^-$), (${\bar D}^0 K^+ K^-$, $D^0 K^+ K^-$), and (${\bar D}^0 K^0
{\bar K}^0$, $D^0 K^0 {\bar K}^0$).  The CKM matrix elements of these
decays are the same as in the corresponding two-body decay modes, and
will therefore exhibit very similar time-dependent CP asymmetries.

The decay amplitude of $\bs ({\bsbar}) \to PPP$ receives several
different contributions, both resonant and non-resonant.  In the
following, we perform a time-dependent Dalitz-plot analysis of the
three-body decays. This permits the measurement of each of the
contributing amplitudes, as well as their relative phases. As we will
see below, the Dalitz-plot analysis reduces the ambiguity in the
measurement of $\gamma$ and $2\beta_s$ compared to the corresponding
two-body decays.  We also show how this analysis resolves the sign
ambiguity in $\Delta\Gamma_s$.

\subsection{Dalitz-plot analysis}

Here we review certain aspects of the Dalitz-plot analysis. We focus
on the general three-body decay $B \to P_1 P_2 P_3$. We define the
Dalitz-plot variables
\be
s_{12} \equiv (p_1 + p_2)^2 ~,~~ s_{13} \equiv (p_1 + p_3)^2 ~,~~ s_{23} \equiv (p_2 + p_3)^2 ~,
\ee
which are related by the conservation law
\be
s_{12} + s_{13} + s_{23} = m^{2}_{B} + m_1^2 + m_2^2 + m_3^2 ~.
\ee
This shows that there are only two independent variables (below, we
use $s_{12}$ and $s_{13}$).

$B \to P_1 P_2 P_3$ can take place either via intermediate resonances
or non-resonant contributions. A widely-used approximation in the
parametrization of the decay amplitude is the isobar model. In this
model, the individual terms are interpreted as complex production
amplitudes for two-body resonances, and one also includes a term
describing the non-resonant component. The amplitude is then written
as
\be
{\cal A}(s_{12},s_{13}) = \sum_j a_j F_j(s_{12},s_{13}) ~, 
\label{resoamp}
\ee
where the sum is over all decay modes (resonant and non-resonant).
Here, the $a_j$ are the complex coefficients describing the magnitudes
and phases of different decay channels, while the $F_j(s_{12},s_{13})$
contain the strong dynamics. The CP-conjugate amplitude is given by
\be
{\bar {\cal A}}(s_{12},s_{13}) = \sum_j {\bar a_j} {\bar F_j}(s_{13},s_{12}) ~,
\ee
where ${\bar F_j}(s_{13},s_{12}) = F_j(s_{12},s_{13})$.

Now, in the experimental analysis, the $F_j(s_{12},s_{13})$ take
different (known) forms for the various contributions. By performing a
maximum likelihood fit over the entire Dalitz plot, one can obtain the
magnitudes and relative phases of the $a_j$, and similarly for the
${\bar a_j}$.  Thus, the full decay amplitudes can be obtained.

\subsection{\boldmath $\bs ({\bsbar}) \to D^{\pm}_s K^{\mp} \pi^0$}
\label{threebodydspi}

In this subsection we focus specifically on the decay $\bs ({\bsbar})
\to D^{\pm}_s K^{\mp} \pi^0$, and use a modification of the method
elaborated in Ref.~\cite{PSS}. The Dalitz-plot variables are
\be
s^+ \equiv (p_{D_s} + p_{\pi})^2 ~,~~ s^- \equiv (p_K + p_{\pi})^2 ~,~~ s^0 \equiv (p_{D_s} + p_{K})^2 ~. 
\ee
The amplitudes are written as
\be
{\cal A}(s^+,s^-) = \sum_j a_j F_j(s^+,s^-) ~,~~ {\bar {\cal A}}(s^+,s^-) = \sum_j {\bar a_j} {\bar F_j}(s^-,s^+) ~.
\label{resoamp}
\ee
The time-dependent decay rates for the oscillating $\bs(t)$ and
$\bsbar(t)$ mesons, decaying to the same final state $f$, are given by
\bea
\Gamma(\bs(t) \to f) &\! \sim \!& \frac12 e^{-\Gamma_s t} 
\Big[A_{ch}(s^+,s^-) \cosh(\Delta \Gamma_s t /2) - A_{sh}(s^+,s^-) \sinh(\Delta \Gamma_s t /2) 
  \nn \\ 
&& \hskip0.4truein
 +~A_{c}(s^+,s^-) \cos(\Delta m_s t) - A_{s}(s^+,s^-) \sin(\Delta m_s t)\Big] ~, \nn\\
\Gamma(\bsbar(t) \to f) &\! \sim \!& \frac12 e^{-\Gamma_s t} 
\Big[A_{ch}(s^-,s^+) \cosh(\Delta \Gamma_s t /2) - A_{sh}(s^-,s^+) \sinh(\Delta \Gamma_s t /2)
  \nn \\
&& \hskip0.4truein
 -~A_{c}(s^-,s^+) \cos(\Delta m_s t) + A_{s}(s^-,s^+) \sin(\Delta m_s t)\Big] ~.
\label{timedalitz}
\eea
Here
\bea
A_{ch}(s^+,s^-) &=& |{\cal A}(s^+,s^-)|^2 + |{\bar {\cal A}}(s^+,s^-)|^2 ~, \nn \\
A_{c}(s^+,s^-) &=& |{\cal A}(s^+,s^-)|^2 - |{\bar {\cal A}}(s^+,s^-)|^2 ~, \nn \\
A_{sh}(s^+,s^-) &=& 2 {\rm Re} \left( e^{-2 i \beta_s} {\bar {\cal A}}(s^+,s^-) {\cal A}^{\ast}(s^+,s^-) \right) ~,  \nn \\
A_{s}(s^+,s^-) &=&  2 {\rm Im} \big( e^{-2 i \beta_s} {\bar {\cal A}}(s^+,s^-) {\cal A}^{\ast}(s^+,s^-) \big) ~,
\label{cdalitz}
\eea
where ${\cal A}(s^+,s^-)$ and $\bar {\cal A}(s^+,s^-)$ are given in
Eq.~(\ref{resoamp}). 

Now, there are a number of resonances which contribute to $\bs
({\bsbar}) \to D^{\pm}_s K^{\mp} \pi^0$. For illustrative purposes, we
consider just two of them: $D^{\pm}_s K^{\ast \mp}(892)$ and $D^{\ast
  \pm}_s K^{\mp}$. The decays $K^{\ast \pm } \to K^{\pm} \pi^0$ and
$D^{\ast \pm}_s\to D^{\pm}_s \pi^0$ have already been observed: $B(K^{\ast
  \pm }(892) \to K^{\pm} \pi^0) = 50\%$, $B(D^{\ast \pm}_s\to
D^{\pm}_s \pi^0) = (5.8 \pm 0.7) \%$ \cite{pdg10}. $\bs ({\bsbar}) \to
D^{\pm}_s K^{\ast \mp}$ and $\bs ({\bsbar}) \to D^{\ast \pm}_s
K^{\mp}$ are the $\bs ({\bsbar}) \to VP$ equivalents of the decay
discussed in Sec.~\ref{2body}, $\bs ({\bsbar}) \to D^{\pm}_s
K^{\mp}$. The additional ingredient here is that we also consider the
decay products of the $V$, so that we have the full decay chain $\bs
({\bsbar}) \to VP \to PPP$.

For these two resonances, we have
\bea
{\cal A}_{K^*}(\bs\to D_s^- K^{*+} \to D_s^- K^+ \pi^0) &=& a^{K^*}_1 e^{i \gamma} F_{K^*} ~, \nn \\
{\bar {\cal A}}_{K^*}(\bsbar\to D_s^- K^{*+} \to D_s^- K^+ \pi^0) &=& a^{K^*}_2  F_{K^*} ~, \nn \\
{\cal A}_{D_s^*}(\bs\to D_s^{*-} K^+ \to D_s^- K^+ \pi^0) &=& a^{D_s^*}_1 e^{i \gamma} F_{D_s^*} ~, \nn \\
{\bar {\cal A}}_{D_s^*}(\bsbar\to D_s^{*-} K^+ \to D_s^- K^+ \pi^0) &=& a^{D_s^*}_2  F_{D_s^*} ~.
\label{decayamp}
\eea
Including both resonances, the amplitudes of $\bs (\bsbar) \to
D_s^- K^+ \pi^0$ are
\bea
{\cal A} &=& e^{i \gamma} ( a^{K^*}_1 F_{K^*} + a^{D_s^*}_1 F_{D_s^*}) ~, \nn \\
{\bar{\cal A}} &=& ( a^{K^*}_2 F_{K^*} + a^{D_s^*}_2 F_{D_s^*}) ~.
\eea

With these amplitudes, $A_{ch}$, $A_{c}$, $A_{sh}$ and $A_s$ [Eq.~(\ref{cdalitz})] take the forms
\bea
A^{D_s K \pi}_{ch} &=& \left(|a^{K^*}_1|^2 + |a^{K^*}_2|^2 \right) |F_{K^*}|^2 + \left( |a^{D_s^*}_1|^2 + |a^{D_s^*}_2|^2 \right) |F_{D_s^*}|^2 \nn \\ 
&& +~2 {\rm Re} \left( (a^{K^*}_1 F_{K^*})^*  (a^{D_s^*}_1 F_{D_s^*}) \right) 
+ 2 {\rm Re} \left( (a^{K^*}_2 F_{K^*})^* (a^{D_s^*}_2 F_{D_s^*}) \right) ~, \nn\\
A^{D_s K \pi}_{c} &=& \left(|a^{K^*}_1|^2 - |a^{K^*}_2|^2 \right) |F_{K^*}|^2 + \left( |a^{D_s^*}_1|^2 - |a^{D_s^*}_2|^2 \right) |F_{D_s^*}|^2 \nn \\
&& +~2 {\rm Re} \left( (a^{K^*}_1 F_{K^*})^* (a^{D_s^*}_1 F_{D_s^*}) \right) 
- 2 {\rm Re} \left( (a^{K^*}_2 F_{K^*})^* (a^{D_s^*}_2 F_{D_s^*}) \right) ~, \nn \\
A^{D_s K \pi}_{sh} &=& \cos(2\beta_s + \gamma + \delta_{K^*}) |a^{K^*}_1| |a^{K^*}_2| |F_{K^*}|^2 + \cos(2\beta_s + 
\gamma + \delta_{D_s^*}) |a^{D_s^*}_1|
  |a^{D_s^*}_2| |F_{D_s^*}|^2 \nn \\
&& +~\cos(2\beta_s + \gamma + \delta) |a^{K^*}_1| |a^{D_s^*}_2|  {\rm Re} \left(F^{\ast}_{K^*} F_{D_s^*}\right) \nn\\
&& -~\sin(2\beta_s + \gamma + \delta) |a^{K^*}_1| |a^{D_s^*}_2|  {\rm Im} \left(F^{\ast}_{K^*} F_{D_s^*}\right) \nn \\
&& +~{\rm Re} \left[ e^{- i (2\beta_s + \gamma) } (a^{D_s^*}_1 F_{D_s^*})^* (a^{K^*}_2  F_{K^*}) )\right]~, \nn \\
A^{D_s K \pi}_s &=& -\sin(2\beta_s + \gamma + \delta_{K^*}) |a^{K^*}_1| |a^{K^*}_2| |F_{K^*}|^2 - \sin(2\beta_s + \gamma + 
\delta_{D_s^*}) |a^{D_s^*}_1| 
  |a^{D_s^*}_2| |F_{D_s^*}|^2 \nn \\
&&  -~\sin(2\beta_s + \gamma + \delta) |a^{K^*}_1| |a^{D_s^*}_2|  {\rm Re} \left(F^{\ast}_{K^*} F_{D_s^*}\right) \nn\\
&& +~\cos(2\beta_s + \gamma + \delta) 
|a^{K^*}_1| |a^{D_s^*}_2|  {\rm Im} \left(F^{\ast}_{K^*} F_{D_s^*}\right) \nn \\
&& +~{\rm Im} \left[ e^{- i (2\beta_s + \gamma) } (a^{D_s^*}_1 F_{D_s^*})^* (a^{K^*}_2  F_{K^*}) )\right] ~,
\label{asydspik}
\eea    
with $\delta_{{K^*}} = - \arg \left( (a^{K^*}_1)^* a^{{K^*}}_2
\right)$, $\delta_{{D_s^*}} = - \arg \left( (a^{D_s^*}_1)^*
a^{{D_s^*}}_2 \right)$, and $\delta = - \arg \left( (a^{K^*}_1)^*
a^{D_s^*}_2 \right)$.

Above, in the discussion of the time-independent Dalitz-plot analysis,
we noted that the magnitudes and relative phases of the $a_j$ can be
obtained from a maximum likelihood fit over the entire Dalitz plot,
given assumed forms for the $F_j$'s. The same holds true for the
time-dependent Dalitz-plot analysis -- the magnitudes and relative
phases of the contributing resonances, i.e.\ $a^{K^*}_1$, $a^{K^*}_2$,
$a^{D_s^*}_1$, and $a^{D_s^*}_2$, can all be obtained. Indeed, such
an analysis has been performed by the Babar and Belle collaborations
for the decay $\bd(t) \to \ks \pi^+ \pi^-$ \cite{timedepDalitz}. In
particular, all the coefficients that multiply the $F^{\ast}_i F_j$
[Eq.~(\ref{asydspik})] bilinears can be obtained from a maximum
likelihood fit to the corresponding Dalitz plot PDFs.

This permits the extraction of the weak phases. For example, we can
extract $2\beta_s + \gamma + \delta$ without any ambiguity from the
third and fourth terms of $A^{D_s K \pi}_s$. In a similar manner, the
time-dependent Dalitz-plot analysis of $\bs ({\bsbar}) \to D^+_s K^-
\pi^0$ allows the extraction of the phase $2\beta_s + \gamma -
\delta$. The combination of these two results yields $2\beta_s +
\gamma$ and $\delta$ with a twofold ambiguity. And if factorization is
imposed, the discrete ambiguity is removed entirely (only the solution
with $\delta \simeq 0$ is kept). The key point here is that we do not
use $A^{D_s K \pi}_{sh}$ at all. As a consequence, there is no
discrete ambiguity due to the sign ambiguity of $\Delta \Gamma_s$ [see
  the discussion following Eq.~(\ref{cdalitz})].  This is to be
contrasted with two-body decays. There $2\beta_s + \gamma$ can also be
obtained with a twofold ambiguity. However, because ${\cal
  A}_{\Delta\Gamma}$ and ${\bar{\cal A}}_{\Delta\Gamma}$ are used
[Eq.~(\ref{2bodyresults})], there is an additional discrete ambiguity
due to the unknown sign of $\Delta\Gamma_s$.

We note that one can extract different trigonometric functions such as
$|\sin(2\beta_s + \gamma + \delta)|$, $|\cos(2\beta_s + \gamma +
\delta)|$, $|\cos(2\beta_s + \gamma + \delta_i)|$, {\rm etc.}, from
$A^{D_s K \pi}_{sh}$ [Eq.~(\ref{asydspik})]. Due to the sign ambiguity
of $\Delta \Gamma_s$, which can be viewed as the sign ambiguity in
$A^{D_s K \pi}_{sh}$, the sign of these trigonometric functions cannot
be determined.  Depending on the sign of $\Delta\Gamma_s$, their sign
could be positive or negative. Therefore, we can determine the sign of
$\Delta\Gamma_s$ if we are able to fix the sign of these trigonometric
functions. Now, the functions $\sin(2\beta_s + \gamma + \delta)$ and
$\cos(2\beta_s + \gamma + \delta)$ can be extracted without ambiguity
from $A^{D_s K \pi}_s$, which fixes the sign of $\Delta \Gamma_s$ and
hence removes the discrete ambiguity in $A^{D_s K\pi}_{sh}$. Note that
this can be done without measuring $\phi_s$.  This method can
therefore be used to determine the sign of $\cos\phi_s$.

In the above, we have concentrated on the decay $\bs ({\bsbar}) \to
D^{\pm}_s K^{\mp} \pi^0$. However, any of the decay pairs discussed in
Sec.~\ref{threebody} can be used. All that is necessary is that there
be at least two resonances contributing to the decay. We therefore see
that, by using such three-body decays, one can obtain $2\beta_s +
\gamma$ (with a twofold ambiguity if factorization is not assumed), as
well as resolve the sign ambiguity in $\Delta \Gamma_s$. The
resolution of the $\Delta\Gamma_s$ sign ambiguity determines the sign
of $\cos\phi_s$. The precise knowledge of $\gamma$ from other
measurements allows one to obtain $2\beta_s$ with a twofold ambiguity
(since $2\beta_s + \gamma$ can itself be extracted with a twofold
ambiguity), which can be compared with the measurement of $2\beta_s$
from $\Bsdecay$ [Eq.~(\ref{cdfphi})].

Still, it is preferable to have a method that allows the direct
determination of $2\beta_s$ and $\gamma$ individually. This can be
done by measuring the decay $\bs (\bsbar) \to D^0_{CP} K {\bar K}$,
which is discussed in the next subsection.

\subsection{\boldmath $\bs(\bsbar) \to D^0_{CP} K {\bar K}$}
\label{threebodydkk}

In Sec.~\ref{twobodydphi} we discussed the two-body decays $\bs
(\bsbar) \to D \phi$ ($D = D^0,\bar D^0, D^0_{CP}$), and showed that
it is possible to extract $2\beta_s$ and $2\gamma$ with a twofold
ambiguity due to the unknown sign of $\Delta\Gamma_s$.  The
time-dependent Dalitz-plot analysis of $\bs (\bsbar)\to D^0 K {\bar
  K}, {\bar D^0} K {\bar K}$ is similar to that of the previous
subsection, with the intermediate resonances $\phi(1020)$ or
$f_0(1500)$ decaying to the final state $K {\bar K}$.  In this
subsection we consider in addition the related three-body decays $\bs
(\bsbar)\to D^0_{CP} K {\bar K}$, with $D^0_{CP} \equiv 1/\sqrt{2}(D^0
\pm {\bar D^0})$.

$\bs (\bsbar)\to D^0_{CP} K {\bar K}$ receives contributions from
several different intermediate resonances: $\phi(1020)$, $\phi(1680)$,
$f_0(1500)$, $f_0(1710)$, $D^{\ast \pm}_{s_j}$, etc., which follow the
decay chains $\bs (\bsbar) \to D^0_{CP} \phi \to D^0_{CP} K^+ K^-$,
$\bs (\bsbar) \to D^0_{CP} f_0 \to D^0_{CP} K^+ K^-$, $\bs (\bsbar)\to
D^{\ast \pm}_{s_j} K^{\mp} \to D^0_{CP} K^{\pm} K^{\mp}$.  To simplify
our analysis, we consider only the $\phi(1020)$ and $f_0(1500)$
resonances.  The amplitude with an intermediate $\phi$ resonance can
be written as
\bea
\sqrt{2} A_{\phi}(\bs \to D^0_{CP} \phi (\to K^+ K^-)) &=& A(\bs \to D^0 K^+ K^-) + A(\bs \to {\bar D^0} K^+ K^-) ~, \nn \\
\sqrt{2} {\bar A_{\phi}}(\bsbar \to D^0_{CP} \phi (\to K^+ K^-)) &=& A(\bsbar \to D^0 K^+ K^-) + A(\bsbar \to {\bar D^0} K^+ K^-) ~,
\label{cpaphi}
\eea
where 
\bea
A(\bs\to D^0 \phi \to D^0 K^+ K^-) &=& -C^{\phi}_1 e^{i \gamma} F_{\phi} ~, \nn \\
A(\bsbar\to D^0 \phi \to D^0 K^+ K^-) &=& -C^{\phi}_2  F_{\phi} ~, \nn \\
A(\bs\to {\bar D^0} \phi \to D^0 K^+ K^-) &=& C^{\phi}_2  F_{\phi} ~, \nn \\
A(\bsbar\to {\bar D^0} \phi \to D^0 K^+ K^-) &=& C^{\phi}_1 e^{-i \gamma} F_{\phi} ~.
\label{decayamp}
\eea
The amplitude with an intermediate $f_0$ resonance is given by a
similar expression, with the replacement $\phi \to f_0$.  Including
the contributions from these two resonances, the total amplitude can
be written as
\bea
{\cal A}(\bs \to D^0_{CP} K^+ K^-) &=&  A_{\phi}(\bs \to D^0_{CP} K^+ K^-)  +  A_{f_0}(\bs \to D^0_{CP} K^+ K^-) ~, \nn \\
{\bar{\cal A}}(\bsbar \to D^0_{CP} K^+ K^-) &=& {\bar A_{\phi}}(\bsbar \to D^0_{CP} K^+ K^-)  +  
{\bar A_{f_0}}(\bsbar \to D^0_{CP} K^+ K^-) ~.
\eea

With these, $A^{DKK}_{c}$, $A^{DKK}_{ch}$ and $A^{DKK}_s$ can be
computed similarly to Eq.~(\ref{asydspik}).  First, we have
\bea
A^{DKK}_{c} &=&  \sum_{i=\phi,f_0} \left[ \left(|A_i|^2 - |{\bar A_{i}}|^2\right) + 2 {\rm Re}  \big( A_{\phi} A^{\ast}_{f_0}
- {\bar A_{\phi}} {\bar A^{\ast}_{f_0}} \big) \right] ~, \nn \\
A^{DKK}_{ch} &=& \sum_{i=\phi,f_0} \left[ \left(|A_i|^2 + |{\bar A_{i}}|^2\right) + 2 {\rm Re}  \big( A_{\phi} A^{\ast}_{f_0}
+ {\bar A_{\phi}} {\bar A^{\ast}_{f_0}} \big) \right] ~,
\label{adkkc}
\eea
in which
\bea
{\rm Re} \big( A_{\phi} A^{\ast}_{f_0} - {\bar A_{\phi}} {\bar
  A^{\ast}_{f_0}} \big) &=& |C^{f_0}_2| |C^{\phi}_2| \sin\gamma \Big[ r_{\phi}
  \big\{ \sin\delta_{\phi} {\rm Re}(F_{\phi}
  F^{\ast}_{f_0}) + \cos\delta_{\phi} {\rm Im}(F_{\phi}
  F^{\ast}_{f_0})\big\} \nn \\
&& \hskip0.75truein + ~r_{f_0} \big\{ \sin\delta_{f_0} {\rm Re}(F_{\phi}
F^{\ast}_{f_0}) - \cos\delta_{f_0} {\rm Im}(F_{\phi}
F^{\ast}_{f_0})\big\}\Big] ~, \nn\\ 
{\rm Re} \big( A_{\phi} A^{\ast}_{f_0} +
{\bar A_{\phi}} {\bar A^{\ast}_{f_0}} \big) &=&  |C^{f_0}_2|
|C^{\phi}_2| \Big[ {\rm Re}(F_{\phi} F^{\ast}_{f_0}) - r_{\phi}~
  \cos\gamma \big\{ \cos\delta_{\phi} {\rm Re}(F_{\phi}
  F^{\ast}_{f_0}) - \sin\delta_{\phi} {\rm Im}(F_{\phi}
  F^{\ast}_{f_0})\big\} \nn \\
&& \hskip0.75truein - ~r_{f_0} \cos\gamma \big\{ \cos\delta_{f_0} {\rm Re}(F_{\phi}
F^{\ast}_{f_0}) + \sin\delta_{f_0} {\rm Im}(F_{\phi}
F^{\ast}_{f_0})\big\} \nn \\ 
&& \hskip0.75truein + ~r_{\phi} r_{f_0}
\big\{\cos(\delta_{\phi} - \delta_{f_0}) {\rm Re}(F_{\phi}
F^{\ast}_{f_0}) - \sin(\delta_{\phi} - \delta_{f_0}) {\rm Im}(F_{\phi}
F^{\ast}_{f_0}) \big\}\Big] ~,
\label{reacp}
\eea
where $r_i = |C^{i}_1|/|C^{i}_2|$ and $\delta_i =
\arg(C^{i}_1/C^{i}_2)$ ($i=\phi,f_0$). Using Eq.~(\ref{reacp}) in
Eq.~(\ref{timedalitz}), a maximum likelihood fit to the Dalitz-plot
PDFs allows one to extract
\bea
 |C^{f_0}_2| |C^{\phi}_2| r_{i} \cos\gamma \ \cos\delta_{i} &\equiv& \sigma^i_c ~,\nn \\
 |C^{f_0}_2| |C^{\phi}_2| r_{i} \sin\gamma \ \cos\delta_{i} &\equiv& \sigma^i_s ~.
\eea
This gives the ratio 
\be
\frac{\sigma^i_s}{\sigma^i_c} = \tan\gamma ~.
\label{tangamma1}
\ee
Since the hadronic uncertainties cancel in the ratio, it yields a
theoretically clean determination of the angle $\gamma$ with a twofold
ambiguity, even without the knowledge of the strong phases.

Second, we have
\beq
A^{DKK}_s = {\rm Im}\left[ e^{-2i\beta_s} {\cal A}^{\ast} {\bar{\cal A}}\right] = {\rm Im} \left[ 
e^{-2i\beta_s} \big( A^{\ast}_{\phi} \bar A_{\phi} + 
A^{\ast}_{\phi} \bar A_{f_0} + A^{\ast}_{f_0} \bar A_{\phi} + A^{\ast}_{f_0} \bar A_{f_0} \big) \right] ~.
\eeq
The first and fourth terms of $A^{DKK}_s$ are given by 
\beq
{\rm Im}\left[e^{-2i\beta_s} A^{\ast}_{i} \bar A_{i}\right] = \frac{1}{2} {\rm
  Im}\left[ e^{-2i\beta_s}|C^{i}_2|^2 |F_i|^2 \big( 1 + r^2_i
  e^{-2i\gamma} + r_i (e^{i(\delta_{\phi}-\gamma)} +
  e^{-i(\delta_{\phi}+\gamma)})\big)\right] ~,
\label{im1}
\eeq
which allows the extraction of $\sin2\beta_s$, $\sin(2\beta_s + 2\gamma)$, $\sin(2\beta_s + \gamma -
\delta_{\phi/f_0})$ and $\sin(2\beta_s + \gamma + \delta_{\phi/f_0})$. The $\phi$-$f_0$ interference terms are given by
\bea
{\rm Im} \left[ e^{-2i\beta_s}\big(A^{\ast}_{\phi} \bar A_{f_0} +
  A^{\ast}_{f_0} \bar A_{\phi}\big)\right] &=& \frac{1}{2} {\rm Im} \Big[
  e^{-2i\beta_s} |C^{\phi}_2| |C^{f_0}_2| \Big\{ - \left(
  F^{\ast}_{\phi} F_{f_0} + F^{\ast}_{f_0} F_{\phi} \right) \nn
  \\
&& \hskip0.25truein +~r_{\phi} \left(e^{-i(\gamma + \delta_{\phi})} F^{\ast}_{\phi}
 F_{f_0} + e^{- i(\gamma - \delta_{\phi})} F^{\ast}_{f_0}
 F_{\phi}\right) \nn \\
&& \hskip0.25truein +~r_{f_0} \left(e^{- i(\gamma-\delta_{f_0})} F^{\ast}_{\phi}
 F_{f_0} + e^{- i(\gamma + \delta_{f_0})} F^{\ast}_{f_0}
 F_{\phi}\right) \nn \\
&& \hskip0.25truein - ~r_{\phi} r_{f_0}\left(e^{- i(2\gamma + \delta_{\phi} -
   \delta_{f_0})} F^{\ast}_{\phi} F_{f_0} + e^{- i(2\gamma -
   \delta_{\phi}+\delta_{f_0})} F^{\ast}_{f_0} F_{\phi}\right)
 \Big\}\Big] ~.
\label{im3}
\eea
This yields
\bea
{\rm Im} \left[ e^{-2i\beta_s}\big( A^{\ast}_{\phi} \bar A_{f_0} +
  A^{\ast}_{f_0} \bar A_{\phi} \big) \right] 
&=& \frac{1}{2} |C^{\phi}_2| |C^{f_0}_2|\Big[ \sin2\beta_s \left\{ {\rm Re} (F^{\ast}_{\phi}
F_{f_0}) + {\rm Re} (F^{\ast}_{f_0} F_{\phi}) \right\}  \nn \\
&& \hskip -20pt -~r_{\phi} \Big\{ \sin(2\beta_s +
\gamma+\delta_{\phi}){\rm Re} \big(F^{\ast}_{\phi} F_{f_0}\big) -
\cos(2\beta_s + \gamma+\delta_{\phi}) {\rm Im} \big(F^{\ast}_{\phi}
F_{f_0}\big) \nn \\
&& \hskip -20pt  +~\sin(2\beta_s + \gamma-\delta_{\phi}) {\rm Re}
 \big(F^{\ast}_{f_0} F_{\phi}\big) - \cos(2\beta_s +
 \gamma-\delta_{\phi}) {\rm Im}\big(F^{\ast}_{f_0} F_{\phi}\big)
 \Big\} \nn \\
&& \hskip -20pt -~r_{f_0} \Big\{\sin(2\beta_s + \gamma+\delta_{f_0})
 {\rm Re} \big(F^{\ast}_{f_0} F_{\phi}\big) -
\cos(2\beta_s + \gamma+\delta_{f_0}) {\rm Im} \big(F^{\ast}_{f_0}
F_{\phi}\big) \nn \\
&& \hskip -20pt  +~\sin(2\beta_s + \gamma-\delta_{f_0}) {\rm Re}
 \big(F^{\ast}_{\phi} F_{f_0}\big) - \cos(2\beta_s +
 \gamma-\delta_{f_0}) {\rm Im}\big(F^{\ast}_{\phi} F_{f_0}\big)\Big\} \nn\\
&& \hskip -20pt 
 +~r_{\phi}\, r_{f_0} \Big\{\sin(\delta_{\phi}-\delta_{f_0}+ 2\beta_s + 2\gamma) {\rm Re}
 \big(F^{\ast}_{\phi} F_{f_0}\big) - \cos(\delta_{\phi}-\delta_{f_0}+
 2\beta_s + 2\gamma) {\rm Im}\big(F^{\ast}_{\phi} F_{f_0}\big) \nn \\
&& \hskip -20pt +~\sin(\delta_{f_0}-\delta_{\phi}+ 2\beta_s +
 2\gamma) {\rm Re} \big(F^{\ast}_{f_0} F_{\phi}\big) -
 \cos(\delta_{f_0}-\delta_{\phi}+ 2\beta_s + 2\gamma) {\rm
   Im}\big(F^{\ast}_{f_0} F_{\phi}\big) \Big\} \Big]~.
\label{im2}
\eea
{}From the above, we can extract 
\bea
- r_i\,\sin(2\beta_s + \gamma \pm \delta_i) &\equiv& S^{i \pm}_{DKK} ~, \hskip 30pt 
 r_i\, \cos(2\beta_s + \gamma \pm \delta_i) \equiv C^{i \pm}_{DKK} ~, \hskip 30pt 
\sin2\beta_s \equiv  S_{DKK} ~, \nn \\
r_{ij} \, \sin(2\beta_s + 2\gamma \pm \delta_{ij}) &\equiv& S^{ij\pm}_{DKK} ~, \hskip 30pt 
-r_{ij}\,\cos(2\beta_s + 2\gamma \pm \delta_{ij}) \equiv C^{ij\pm}_{DKK} ~,
\label{obs:dkk}
\eea
where $r_{ij} \equiv r_i r_j$ and the corresponding $\delta_{ij}
\equiv \delta_i - \delta_j$ ($i,j = \phi, f_0$). It is straightforward
to find expressions for $\tan(2\beta_s + \gamma)$ and $\tan(2\beta_s
+ 2\gamma)$ in terms of the above observables:
\be
\tan(2\beta_s + \gamma) = - \frac{S^{i +}_{DKK} + S^{i -}_{DKK}}{C^{i +}_{DKK} + C^{i -}_{DKK}} ~, \hskip 30pt  
\tan(2\beta_s + 2\gamma) = - \frac{S^{ij +}_{DKK} + S^{ij -}_{DKK}}{C^{ij +}_{DKK} + C^{ij -}_{DKK}} ~. 
\label{tanfun1}
\ee
With these, one can obtain the expression for
$\tan\gamma$ in terms of the extracted observables:
\be
\tan\gamma = \frac{\tan(2\beta_s + 2\gamma ) - \tan(2\beta_s + \gamma)}{1 - \tan(2\beta_s + \gamma) 
\tan(2\beta_s + 2\gamma)} ~.
\label{tangamma}
\ee
This way of getting $\tan\gamma$ uses $A^{DKK}_s$ [see also Eq.~(\ref{tangamma1})].

Combining Eqs.~(\ref{tanfun1}) and (\ref{tangamma}), we obtain
\be
\tan2\beta_s = \frac{\tan(2\beta_s + \gamma) - \tan\gamma}{1 - \tan(2\beta_s + \gamma) \tan\gamma} ~.
\label{tan2betas}
\ee
This determines $2\beta_s$ with the twofold ambiguity $2\beta_s \to
\pi + 2\beta_s$. However, as we note in Eq.~(\ref{obs:dkk}), we can
extract $\sin2\beta_s$ without any sign ambiguity. This determines
$2\beta_s$ with the twofold ambiguity $2\beta_s \to \pi - 2\beta_s$,
which is different from that obtained in $\tan2\beta_s$.  Therefore,
the combined measurements of $\tan2\beta_s$ and $\sin2\beta_s$ allow
us to extract $2\beta_s$ without any ambiguity. The sign ambiguity in
$\Delta\Gamma_s$ can be resolved in a similar way to that discussed in
Sec.~\ref{threebodydspi}.

Above, we discussed the interference between the two resonance states
$\phi(1020)$ and $f_0(1500)$. However, the analysis would hold equally
for the interference between any two resonances decaying to the same
final state.  Similar information can also be obtained from the
time-dependent Dalitz-plot analysis of $\bs(\bsbar) \to D^0_{CP} K^0
{\bar K^0}$.

\section{\boldmath Extraction of $\phi_s$ and $|\Gamma^s_{12}|$}
\label{gamma12exp}

In the previous section(s) we examined methods for extracting the CP
phase $2\beta_s$ using various two- and three-body decays. The idea is
that if a nonzero value of $2\beta_s$ is found, this will be clear
evidence of new physics in $\bs$-$\bsbar$ mixing.  In addition, if
such a value of $2\beta_s$ is obtained, we will want to know its exact
value in order to ascertain which different models of NP could
generate such mixing. To this end, the best method will be that for
which the discrete ambiguity in $2\beta_s$ is minimized. However,
there is one question which has not yet been addressed: if NP in the
mixing is found, does it contribute to $\Gamma^s_{12}$ in addition to
$M^s_{12}$?

In order to answer this question, $\phi_s$ and $|\Gamma^s_{12}|$ must
be measured and their values compared with those predicted by the SM
[Eq.~(\ref{DGSSM})].  As pointed out by several authors, the
possibility of NP in $\Gamma^s_{12}$ cannot be ruled out, and NP can
enhance $\Gamma^s_{12}$ significantly above the SM prediction, even
including the error bars. In order to measure or extract both $\phi_s$
and $|\Gamma^s_{12}|$, we need (at least) two observables which are
sensitive to NP effects in $\Gamma_{12}^s$. We choose $\Delta\Gamma_s$
and the semileptonic asymmetry $a_{sl}^s$.  As we show below, the
precise measurement of these two observables can be used to extract
$\phi_s$ and $|\Gamma^s_{12}|$. 

The expression for $\Delta\Gamma_s$ is given in
Eq.~(\ref{eq:mix_para}); the semileptonic asymmetry is defined as
\beq
a_{sl}^s = {\rm Im} \left[\frac{\Gamma_{12}^s}{M_{12}^s}\right] = \frac{2 |\Gamma_{12}^s|}{\Delta M_s} \sin\phi_s ~. 
\label{asl}
\eeq
Combining Eqs.~(\ref{eq:mix_para}) and (\ref{asl}) we obtain
\bea
\tan\phi_s &=& \frac{a_{sl}^s\,\Delta M_s}{\Delta\Gamma_s}~, \nonumber \\
|\Gamma_{12}^s| &=& \frac{\sqrt{{\Delta\Gamma_s}^2 + {a_{sl}^s}^2\,{\Delta M_s}^2}}{2} ~.  
\label{phis}
\eea
Now, $\Delta M_s$ is known very precisely -- $\Delta M_s = 17.77 \pm
0.12$ \cite{bsmixing,hfag10} -- and so the precise measurements of
$a_{sl}^s$ and $\Delta\Gamma_s$ (without sign ambiguity) allow one to
extract $\phi_s$ without any ambiguity\footnote{Knowledge of
  $\tan\phi_s$ gives $\phi_s$ with a twofold ambiguity, $\phi_s
  \leftrightarrow \pi + \phi_s$. However, $a_{sl}^s$ determines
  $\sin\phi_s$, which allows one to differentiate $\phi_s$ and $\pi +
  \phi_s$.}.  This then determines $|\Gamma_{12}^s|$.

CDF and D\O\ have measured $a_{sl}^s$ directly and the average of
their measurements is given by \cite{hfag10}
\beq
a_{sl}^s = -0.0115 \pm 0.0061 ~.
\eeq
If we take $|\Delta\Gamma_s| = 0.075 \pm 0.04$, as given by CDF
\cite{cdfnew}, we obtain
\beq
\tan\phi_s = -2.72\pm 2.05 ~, \hskip 2cm |\Gamma_{12}^s|= 0.11 \pm 0.05 \hskip 2pt ps^{-1} ~,    
\eeq
while $|\Delta\Gamma_s| = 0.163^{+0.065}_{-0.064}$, as given by
D\O\ \cite{d0new}, yields
\beq
\tan\phi_s = -1.25 \pm 0.83  ~, \hskip 2cm |\Gamma_{12}^s|= 0.13 \pm 0.05 \hskip 2pt ps^{-1} ~.    
\eeq
Although  $|\Gamma_{12}^s|$ and $\phi_s$ can significantly deviate
from their SM predictions [Eq.~(\ref{DGSSM})], both of them are consistent 
with the SM within the error bar. Note that in the above
numerical analysis we do not consider the negative solution for
$\Delta\Gamma_s$, which introduces a sign ambiguity in the extraction
of $\phi_s$. It is clear that improved measurements of both $a_{sl}^s$
and $\Delta\Gamma_s$ are essential in order to understand the
underlying physics of $\bs$-$\bsbar$ mixing and the width difference.

In this paper we have focused on methods for measuring
$\Delta\Gamma_s$ using three-body decays.  In practice, this will be
carried out as follows. For definitiveness, consider the decays $\bs
(\bsbar) \to D^0_{CP} K^+ K^-$. Generalising Eq.~(\ref{timedalitz}) to
$\bs (\bsbar) \to D^0_{CP} K^+ K^-$, the time-dependent untagged
differential decay distribution is given by
\bea
\Gamma_{untagged}(D^0_{CP} K^+ K^-,t) &=& \frac{d^2\Gamma(\bs\to D^0_{CP} K^+ K^-)}{ds^+ ds^-} + \frac{d^2\Gamma(\bsbar\to D^0_{CP} K^+ K^-)}{ds^+ ds^-} \nn \\
 &\equiv& e^{-\Gamma_s t}\left[ A^{DKK}_{ch} \cosh(\Delta \Gamma_s t /2)
 + A^{DKK}_{sh}  \sinh(\Delta \Gamma_s t /2) \right] ~,
\label{fitgamma1}
\eea
where $A^{DKK}_{ch}$ and $A^{DKK}_{sh}$ are defined in
Eq.~(\ref{cdalitz}). Neglecting terms of order
$(\Delta\Gamma_s/\Gamma^2_s)^2$ and higher, the time-integrated
differential untagged decay distribution is given by
\bea
\int^{\infty}_0{dt\,\Gamma_{untagged}(D^0_{CP} K^+ K^-,t)} &=&
 \frac{1}{4 \Gamma_s}\left[A^{DKK}_{ch} + 2 A^{DKK}_{sh} \frac{\Delta\Gamma_s}{\Gamma_s}\right] ~.
\label{fitgamma2}
\eea
For a single resonance, say $\phi$,
\bea
A^{DKK}_{ch} &=& A^2_{\phi} + {\bar A}^2_{\phi}, \nn \\ 
A^{DKK}_{sh} &=& {\rm Re}\left[e^{-2i\beta_s} |C^{\phi}_2|^2 |F_{\phi}|^2 \big\{1 + {r_{\phi}}^2 e^{-2i\gamma} + r_{\phi}
(e^{-i(\gamma + \delta_{\phi})} + e^{-i(\gamma - \delta_{\phi})})\big\}\right] ~.
\label{Ashdef}
\eea
As discussed in the previous section, $A^{DKK}_{ch}$ is fully known
from the CP-averaged branching fraction of the intermediate resonance
$\phi$. Once we have enough precision, a fit to the distribution given
by Eq.~(\ref{fitgamma1}) or (\ref{fitgamma2}) allows one to obtain
$\Delta\Gamma_s$ and the various coefficients of $|F_{\phi}|^2$ (which
yields $2\beta_s$). Such a fit will not allow the determination of the
sign of $\Delta\Gamma_s$ or $\cos\phi_s$, but Eq.~(\ref{phis}) can
still be used to obtain $\phi_s$ (with a twofold ambiguity) and
$|\Gamma_{12}^s|$.

However, the above fit, though possible, is made difficult due to the
requirement of having to simultaneously extract $\Delta\Gamma_s$ and
the components of $A^{DKK}_{ch}$.  Given this, we would rather propose
an alternative procedure. Referring again to Eq.~(\ref{timedalitz}),
the time-dependent tagged differential decay distribution is given by
\bea
\Gamma_{tagged}(D^0_{CP} K^+ K^-,t) &=& \frac{d^2\Gamma(\bs\to D^0_{CP} K^+ K^-)}{ds^+ ds^-} - \frac{d^2\Gamma(\bsbar\to D^0_{CP} K^+ K^-)}{ds^+ ds^-} \nn \\
 &\equiv& e^{-\Gamma_s t}\left[ A^{DKK}_{c} \cos(\Delta m_s t /2)
 - A^{DKK}_{s}  \sin(\Delta m_s t /2) \right] ~,
\label{fittag1}
\eea
where $A^{DKK}_{c}$ and $A^{DKK}_{s}$ are defined in
Eqs.~(\ref{cdalitz}), (\ref{adkkc}) and (\ref{im2}).  {}From a fit to
the above distribution, one can extract only the coefficients of
different bilinears in $A^{DKK}_{s}$ and $A^{DKK}_{c}$, since $\Delta
m_s$ is known. Thus, this fit straightforwardly gives information
regarding $A^{DKK}_{c}$ and $A^{DKK}_{s}$.  As discussed in the
previous section, from $A^{DKK}_s$ alone we can extract $2\beta_s$ and
$\gamma + 2\beta_s \pm \delta_{\phi}$ without any ambiguity, and
$\gamma$ with the ambiguity $[\gamma, \pi+\gamma]$. This permits the
reconstruction of $A^{DKK}_{sh}$ [Eq.~(\ref{Ashdef})]. That is, all
the coefficients of $|F_{\phi}|^2$ in $A^{DKK}_{sh}$ can be obtained
from a fit to Eq.~(\ref{fittag1}). With this knowledge, there is only
one unknown in Eq.~(\ref{fitgamma1}) or (\ref{fitgamma2}) --
$\Delta\Gamma_s$ -- and this can be determined by a fit. This may be a
somewhat simpler procedure. Once we are able to measure
$\Delta\Gamma_s$, then, along with $a_{sl}^s$ and $\Delta M_s$,
Eq.~(\ref{phis}) can be used to obtain the CP phase $\phi_s$ and
$|\Gamma_{12}^s|$.

The above analysis is also applicable to the decays $\bs ({\bsbar})
\to D^{\pm}_s K^{\mp} \pi^0$. However, as was discussed in
Sec.~\ref{threebodydspi}, for such decays all the trigonometric
functions in $A^{D_s K\pi}_{sh}$ are not fully known -- the only known
functions are those appearing as the coefficients of ${\rm Im}(F_i
F^{\ast}_j)$ or ${\rm Re}(F_i F^{\ast}_j)$ ($i \ne j$).  Therefore,
for such decays we can use Eq.~(\ref{fitgamma1}) to fit
$\Delta\Gamma_s$, but we need at least two interfering resonances, and
only the terms proportional to ${\rm Im}(F_i F^{\ast}_j)$ or ${\rm
  Re}(F_i F^{\ast}_j)$ are useful [see $A^{D_s K \pi}_{ch}$ in
  Eq.~(\ref{asydspik})].

\section{Conclusions}
\label{conclusions}

It is well known that the weak phase of $\bs$-$\bsbar$ mixing is very
small in the SM: $2\beta_s \simeq 0$. If this quantity is measured to
be significantly different from zero, this is a smoking-gun signal of
new physics (NP). However, in general we would like more information
from such a measurement. For instance, in order to distinguish among
potential NP models, it is important to have an unambiguous
determination of $2\beta_s$. Similarly, although the width difference
$\Delta\Gamma_s$ between the two $B_s$ mass eigenstates is positive in
the SM, it can take either sign in the presence of NP.  Ideally, a
method probing $\bs$-$\bsbar$ mixing which relies on a nonzero
$\Delta\Gamma_s$ should be able to remove its sign ambiguity.
Finally, although it is usually assumed that NP contributes only to
$M^s_{12}$, it has been shown that NP contributions to $\Gamma^s_{12}$
can also be important. In order to explore this possibility, it is
necessary to measure the CP phase $\phi_s$ and $|\Gamma^s_{12}|$.

In this paper, we examine a variety of methods of measuring
$\bs$-$\bsbar$ mixing with an eye to addressing the above issues. We
look at two- and three-body $B_s$ decays with ${\bar b} \to {\bar c} u
{\bar s}$ and ${\bar b} \to {\bar u} c {\bar s}$ transitions,
concentrating on those final states which are accessible to both $\bs$
and $\bsbar$ mesons (so that there is indirect CP violation). The
time-dependent decay rates include both $\Delta m_s t$ and
$\Delta\Gamma_s t$ terms.

We begin with a review of $\bs (\bsbar) \to D^{\pm}_s K^{\mp}$
decays. Considering sizeable $\Delta\Gamma_s$, we find that this
method allows the extraction of $2\beta_s + \gamma$ with a fourfold
ambiguity. We then turn to $\bs (\bsbar) \to D^0_{CP} \phi$ decays,
where $D^0_{CP}$ is a CP eigenstate. Here we find that $2\beta_s$ and
$2\gamma$ can each be determined up to a twofold ambiguity. Here, the
ambiguity is due to the unknown sign of $\Delta\Gamma_s$.  Therefore,
once we are able to resolve the sign ambiguity in $\Delta\Gamma_s$ by
some other means, the $\bs (\bsbar) \to D^0_{CP} \phi$ decays are
useful to measure $2\beta_s$ and $2\gamma$ without any ambiguity.

In order to resolve the sign ambiguity in $\Delta\Gamma_s$, and to
reduce the discrete ambiguity in the measurement of $2\beta_s$ and
$\gamma$, it is necessary to turn to Dalitz-plot analyses of
three-body decays. We begin with $\bs ({\bsbar}) \to D^{\pm}_s K^{\mp}
\pi^0$. We find that it is possible to obtain $2\beta_s + \gamma$ with
a twofold ambiguity, and to remove the sign ambiguity in
$\Delta\Gamma_s$ (for this, it is not necessary to determine
$\phi_s$). The most promising method involves the decays $\bs(\bsbar)
\to D^0_{CP} K {\bar K}$, in which all issues can be resolved. We find
that $2\beta_s$ can be obtained without any ambiguity, and at the same
we can remove the sign ambiguity in $\Delta\Gamma_s$.  In addition,
$\gamma$ can be determined up to a twofold ambiguity.

Finally, all such decays allow the extraction of $\Delta\Gamma_s$
directly from a fit to the time-dependent untagged differential decay
rate distribution. Given the measurements of $\Delta M_s$, the
semileptonic asymmetry $a_{sl}^s$, and $\Delta\Gamma_s$, the CP phase
$\phi_s$ and $|\Gamma^s_{12}|$ can be obtained. In the case of
three-body decays the coefficients of $\sinh[\Delta\Gamma_s t/2]$ and
$\cosh[\Delta\Gamma_s t/2]$ can be found, either fully or partially,
from a fit to the time-dependent tagged differential decay rate
distribution.  (Of the several three-body decays that we discuss, the
decays $\bs(\bsbar) \to D^0_{CP} K {\bar K}$ are the most promising,
since in such decays these coefficients can be fully reconstructed
from this fit.) Therefore, in three-body decays the only unknown in
the untagged rate distribution is $\Delta\Gamma_s$. This makes the fit
considerably simpler.

\begin{acknowledgements}
SN thanks Ulrich Nierste and Thorsten Feldmann for useful discussions. This work is
financially supported by NSERC of Canada.
\end{acknowledgements}


\begin{thebibliography}{99}

\bibitem{kundu_nandi}
S.~Nandi and A.~Kundu,
  arXiv:hep-ph/0407061;
S.~Mishima and T.~Yoshikawa,
  Phys.\ Rev.\  D {\bf 70}, 094024 (2004)
  [arXiv:hep-ph/0408090];
C.~S.~Kim, S.~Oh and C.~Yu,
  Phys.\ Rev.\  D {\bf 72}, 074005 (2005)
  [arXiv:hep-ph/0505060].

\bibitem{piKupdate} In the latest update of the $\pi K$ puzzle, it was
  seen that, although NP was hinted at in $\btopik$ decays, it could
  be argued that the SM can explain the data, see 
S.~Baek, C.~W.~Chiang and D.~London,
  Phys.\ Lett.\  B {\bf 675}, 59 (2009)
  [arXiv:0903.3086 [hep-ph]].

\bibitem{btos}
 H.~Y.~Cheng, C.~K.~Chua and A.~Soni,
  Phys.\ Rev.\  D {\bf 72}, 094003 (2005)
  [arXiv:hep-ph/0506268];
G.~Buchalla, G.~Hiller, Y.~Nir and G.~Raz,
  JHEP {\bf 0509}, 074 (2005)
  [arXiv:hep-ph/0503151];
E.~Lunghi and A.~Soni,
  JHEP {\bf 0908}, 051 (2009)
  [arXiv:0903.5059 [hep-ph]].

\bibitem{phiK*} B.~Aubert {\it et al.}  [BABAR Collaboration],
  Phys.\ Rev.\ Lett.\  {\bf 91}, 171802 (2003)
  [arXiv:hep-ex/0307026];
  K.~F.~Chen {\it et al.}  [Belle Collaboration],
  Phys.\ Rev.\ Lett.\  {\bf 91}, 201801 (2003)
  [arXiv:hep-ex/0307014].

\bibitem{Belle}
A.~Ishikawa {\it et al.}  [Belle Collaboration],
  Phys.\ Rev.\ Lett.\  {\bf 96}, 251801 (2006)
  [arXiv:hep-ex/0603018];
J.~T.~Wei {\it et al.}  [BELLE Collaboration],
  Phys.\ Rev.\ Lett.\  {\bf 103}, 171801 (2009)
  [arXiv:0904.0770 [hep-ex]].

\bibitem{BaBar}
B.~Aubert {\it et al.}  [BABAR Collaboration],
  Phys.\ Rev.\  D {\bf 73}, 092001 (2006)
  [arXiv:hep-ex/0604007],
  Phys.\ Rev.\  D {\bf 79}, 031102 (2009)
  [arXiv:0804.4412 [hep-ex]].

\bibitem{Dunietz:2000}
  I.~Dunietz, R.~Fleischer and U.~Nierste,
  Phys.\ Rev.\  D {\bf 63}, 114015 (2001)
  [arXiv:hep-ph/0012219].

\bibitem{bsmixing}
V.~M.~Abazov {\it et al.}  [D\O\ Collaboration],
 Phys.\ Rev.\ Lett.\  {\bf 97}, 021802 (2006)
 [arXiv:hep-ex/0603029];
A.~Abulencia {\it et al.}  [CDF - Run II Collaboration],
  Phys.\ Rev.\ Lett.\  {\bf 97}, 062003 (2006)
  [arXiv:hep-ex/0606027].

\bibitem{lenz_uli}
A.~Lenz and U.~Nierste,
  arXiv:1102.4274 [].

\bibitem{CDF} T.~Aaltonen {\it et al.}  [CDF Collaboration],
  Phys.\ Rev.\ Lett.\  {\bf 100}, 161802 (2008)
  [arXiv:0712.2397 [hep-ex]].

\bibitem{D0} V.~M.~Abazov {\it et al.}  [D\O\ Collaboration],
  Phys.\ Rev.\ Lett.\  {\bf 101}, 241801 (2008)
  [arXiv:0802.2255 [hep-ex]].

\bibitem{RPV}
S.~Nandi and J.~P.~Saha,
  Phys.\ Rev.\  D {\bf 74}, 095007 (2006)
  [arXiv:hep-ph/0608341];
A.~Kundu and S.~Nandi,
  Phys.\ Rev.\  D {\bf 78}, 015009 (2008)
  [arXiv:0803.1898 []];
G.~Bhattacharyya, K.~B.~Chatterjee and S.~Nandi,
  Phys.\ Rev.\  D {\bf 78}, 095005 (2008)
  [arXiv:0809.3300 []].

\bibitem{Z'FCNC} V.~Barger, L.~Everett, J.~Jiang, P.~Langacker, T.~Liu and C.~Wagner,
  Phys.\ Rev.\  D {\bf 80}, 055008 (2009)
  [arXiv:0902.4507 [hep-ph]],
  arXiv:0906.3745 [hep-ph].

\bibitem{2HDM} Some aspects of the 2HDM are discussed in
A.~S.~Joshipura and B.~P.~Kodrani,
  Phys.\ Rev.\ D {\bf 81}, 035013 (2010)
  [arXiv:0909.0863 [hep-ph]].
See also
A.~Datta and P.~J.~O'Donnell,
  Phys.\ Rev.\  D {\bf 72}, 113002 (2005)
  [arXiv:hep-ph/0508314];
A.~Datta,
  Phys.\ Rev.\  D {\bf 74}, 014022 (2006)
  [arXiv:hep-ph/0605039].

\bibitem{SUSY}
A.~Datta and S.~Khalil,
  Phys.\ Rev.\ D {\bf 80}, 075006 (2009)
  [arXiv:0905.2105 [hep-ph]].

\bibitem{littleHiggs} M.~Blanke, A.~J.~Buras, S.~Recksiegel
  and C.~Tarantino,
 arXiv:0805.4393 [hep-ph].

\bibitem{fourGen}
A.~Soni, A.~K.~Alok, A.~Giri, R.~Mohanta and S.~Nandi,
  arXiv:0807.1971 [hep-ph];
M.~Bobrowski, A.~Lenz, J.~Riedl and J.~Rohrwild,
  Phys.\ Rev.\  D {\bf 79}, 113006 (2009)
  [arXiv:0902.4883 [hep-ph]];
A.~Soni, A.~K.~Alok, A.~Giri, R.~Mohanta and S.~Nandi,
  Phys.\ Rev.\  D {\bf 82}, 033009 (2010)
  [arXiv:1002.0595 []];
A.~J.~Buras, B.~Duling, T.~Feldmann, T.~Heidsieck, C.~Promberger and S.~Recksiegel,
  JHEP {\bf 1009}, 106 (2010)
  [arXiv:1002.2126 []];

\bibitem{Bspapers} 
J.~K.~Parry and H.~h.~Zhang,
  Nucl.\ Phys.\  B {\bf 802}, 63 (2008)
  [arXiv:0710.5443 [hep-ph]],
B.~Dutta and Y.~Mimura,
  Phys.\ Rev.\  D {\bf 78}, 071702 (2008)
  [arXiv:0805.2988 [hep-ph]],
  Phys.\ Lett.\  B {\bf 677}, 164 (2009)
  [arXiv:0902.0016 [hep-ph]].
J.~h.~Park and M.~Yamaguchi,
  Phys.\ Lett.\  B {\bf 670}, 356 (2009)
  [arXiv:0809.2614 [hep-ph]];
P.~Ko and J.~h.~Park,
  Phys.\ Rev.\  D {\bf 80}, 035019 (2009)
  [arXiv:0809.0705 [hep-ph]];
N.~Kifune, J.~Kubo and A.~Lenz,
Extension
  Phys.\ Rev.\  D {\bf 77} (2008) 076010
  [arXiv:0712.0503 [hep-ph]];
K.~Kawashima, J.~Kubo and A.~Lenz,
  arXiv:0907.2302 [hep-ph];
F.~J.~Botella, G.~C.~Branco and M.~Nebot,
  Phys.\ Rev.\  D {\bf 79}, 096009 (2009)
  [arXiv:0805.3995 [hep-ph]].

\bibitem{NPdecay} C.~W.~Chiang, A.~Datta, M.~Duraisamy, D.~London, M.~Nagashima and A.~Szynkman,
  JHEP {\bf 1004}, 031 (2010)
  [arXiv:0910.2929 [hep-ph]].

\bibitem{cdfnew}
T.~Aaltonen {\it et al.}  [CDF Collaboration],
  arXiv:1112.1726 [hep-ex].

\bibitem{d0new}
V.~M.~Abazov {\it et al.}  [D0 Collaboration],
  arXiv:1109.3166 [hep-ex].


\bibitem{d0dimuonprd}
V.~M.~Abazov {\it et al.}  [D\O\ Collaboration],
  Phys.\ Rev.\  D {\bf 82}, 032001 (2010)
  [arXiv:1005.2757 [hep-ex]].

\bibitem{d0dimuonprl}
 V.~M.~Abazov {\it et al.}  [D\O\ Collaboration],
  Phys.\ Rev.\ Lett.\  {\bf 105}, 081801 (2010)
  [arXiv:1007.0395 [hep-ex]].

\bibitem{dimuon_mixing}
B.~A.~Dobrescu, P.~J.~Fox and A.~Martin,
  Phys.\ Rev.\ Lett.\  {\bf 105}, 041801 (2010)
  [arXiv:1005.4238 []];
C.~H.~Chen, C.~Q.~Geng and W.~Wang,
  JHEP {\bf 1011}, 089 (2010)
  [arXiv:1006.5216 []];
P.~Ko and J.~h.~Park,
  Phys.\ Rev.\  D {\bf 82}, 117701 (2010)
  [arXiv:1006.5821 []];
A.~Lenz {\it et al.},
  Phys.\ Rev.\  D {\bf 83}, 036004 (2011)
  [arXiv:1008.1593 []];
S.~Nandi and A.~Soni,
  Phys.\ Rev.\ D {\bf 83}, 114510 (2011)
  [arXiv:1011.6091 [hep-ph]].

\bibitem{dkn2}
A.~Dighe, A.~Kundu and S.~Nandi,
  Phys.\ Rev.\ D {\bf 82}, 031502 (2010)
  [arXiv:1005.4051 [hep-ph]].

\bibitem{bauer}
C.~W.~Bauer and N.~D.~Dunn,
  Phys.\ Lett.\  B {\bf 696}, 362 (2011)
  [arXiv:1006.1629 []];

\bibitem{dkn1}
A.~Dighe, A.~Kundu and S.~Nandi,
  Phys.\ Rev.\  D {\bf 76}, 054005 (2007)
  [arXiv:0705.4547 []].

\bibitem{NPGamma12} 
N.~G.~Deshpande, X.~G.~He and G.~Valencia,
  arXiv:1006.1682 [hep-ph];
A.~K.~Alok, S.~Baek and D.~London,
  arXiv:1010.1333 [hep-ph].
A.~Datta, M.~Duraisamy, S.~Khalil,
  Phys.\ Rev.\  {\bf D83}, 094501 (2011).
  [arXiv:1011.5979 [hep-ph]].

\bibitem{fleischerdsk}
R.~Fleischer,
  Nucl.\ Phys.\  B {\bf 671}, 459 (2003)
  [arXiv:hep-ph/0304027].

\bibitem{nandi_uli}
S.~Nandi and U.~Nierste,
  Phys.\ Rev.\  D {\bf 77}, 054010 (2008)
  [arXiv:0801.0143 [hep-ph]];
S.~Nandi,
  Nucl.\ Phys.\ Proc.\ Suppl.\  {\bf 209}, 164 (2010).

\bibitem{Gronau:1990ra}
M.~Gronau and D.~London,
  Phys.\ Lett.\  B {\bf 253}, 483 (1991).

\bibitem{fleischerdphi}
R.~Fleischer,
  Phys.\ Lett.\  B {\bf 562}, 234 (2003)
  [arXiv:hep-ph/0301255];
R.~Fleischer,
  Nucl.\ Phys.\  B {\bf 659}, 321 (2003)
  [arXiv:hep-ph/0301256].

\bibitem{Dunietz} I.~Dunietz,
  Phys.\ Rev.\  D {\bf 52}, 3048 (1995)
  [arXiv:hep-ph/9501287].

\bibitem{Ciuchini:2006st}
M.~Ciuchini, M.~Pierini and L.~Silvestrini,
  Phys.\ Lett.\  B {\bf 645}, 201 (2007)
  [arXiv:hep-ph/0602207];
M.~Gronau, D.~Pirjol, A.~Soni and J.~Zupan,
  Phys.\ Rev.\  D {\bf 75}, 014002 (2007)
  [arXiv:hep-ph/0608243].

\bibitem{nicmax}
N.~R.~-L.~Lorier, M.~Imbeault and D.~London,
  Phys.\ Rev.\ D {\bf 84}, 034040 (2011)
  [arXiv:1011.4972 [hep-ph]].
M.~Imbeault, N.~R.~-L.~Lorier and D.~London,
  Phys.\ Rev.\ D {\bf 84}, 034041 (2011)
  [arXiv:1011.4973 [hep-ph]].

\bibitem{PSS} F.~Polci, M.~H.~Schune and A.~Stocchi,
  arXiv:hep-ph/0605129.

\bibitem{pdg10}
K.~Nakamura {\it et al.}  [Particle Data Group],
  J.\ Phys.\ G {\bf 37}, 075021 (2010).

\bibitem{timedepDalitz} B.~Aubert {\it et al.}  [BABAR Collaboration],
  arXiv:0708.2097 [hep-ex];
J.~Dalseno {\it et al.}  [Belle Collaboration],
  Phys.\ Rev.\  D {\bf 79}, 072004 (2009)
  [arXiv:0811.3665 [hep-ex]].
\bibitem{hfag10}
D.~Asner {\it et al.}  [Heavy Flavor Averaging Group Collaboration],
  arXiv:1010.1589 [hep-ex].



\end{thebibliography}
\end{document}